\documentclass[notitlepage,nofootinbib,preprintnumbers,amssymb,superscriptaddress]{revtex4-1}
\usepackage{amsfonts,amssymb,mathtools,graphicx,color,bm,caption,subcaption}
\usepackage{tikz}
\definecolor{ultramarine}{rgb}{0.07, 0.04, 0.56}
\definecolor{cadmiumgreen}{rgb}{0.0, 0.42, 0.24}
\definecolor{indigo(dye)}{rgb}{0.0, 0.25, 0.42}
\usepackage[linktocpage=true,breaklinks]{hyperref}
\hypersetup{
colorlinks=true,
citecolor=ultramarine,
linkcolor=cadmiumgreen,
urlcolor=indigo(dye),
}

\newcommand{\fr}[2]{\frac{#1}{#2}}
\newcommand{\pa}{\partial}

\newcommand{\del}{\delta}

\newcommand{\bra}[1]{\left( #1 \right)}
\newcommand{\brb}[1]{\left[ #1 \right]}
\newcommand{\brc}[1]{\left\{ #1 \right\}}
\newcommand{\be}{\begin{equation%
}}
\newcommand{\ee}{\end{equation}}

\newcommand{\eem}{\end{bmatrix}}

\newcommand{\ep}{\epsilon}

\newcommand{\la}{\lambda}

\newcommand{\mn}{{\mu \nu}}
\newcommand{\mA}{\mathcal{A}}
\newcommand{\mB}{\mathcal{B}}

\newcommand{\mE}{\mathcal{E}}

\newcommand{\mL}{\mathcal{L}}

\begin{document}

\preprint{RUP-20-6, KOBE-COSMO-20-02}

\title{Extended Cuscuton as Dark Energy}

\author{Aya Iyonaga}
\affiliation{Department of Physics, Rikkyo University, Toshima, Tokyo 171-8501, Japan}

\author{Kazufumi Takahashi}
\affiliation{Department of Physics, Kobe University, Kobe 657-8501, Japan}

\author{Tsutomu Kobayashi}
\affiliation{Department of Physics, Rikkyo University, Toshima, Tokyo 171-8501, Japan}

\begin{abstract}
Late-time cosmology in the extended cuscuton theory is studied, in which gravity is modified while one still has no extra dynamical degrees of freedom other than two tensor modes. We present a simple example admitting analytic solutions for the cosmological background evolution that mimics $\Lambda$CDM cosmology. We argue that the extended cuscuton as dark energy can be constrained, like usual scalar-tensor theories, by the growth history of matter density perturbations and the time variation of Newton's constant.
\end{abstract}

\maketitle
\section{Introduction}
General relativity (GR) has been the most successful gravitational theory.
It has the Newtonian and post-Newtonian limit consistent with experiments,
and predicts the existence of gravitational waves and black holes,
which has been directly
confirmed in recent years~\cite{TheLIGOScientific:2016pea,Akiyama:2019cqa}.
Moreover, GR with a cosmological constant can explain the present
accelerated expansion of the Universe.

Due to its simplicity, the cosmological constant has been an appealing candidate
for the origin of the present accelerated expansion.
In order to test this paradigm, it is helpful to compare it with alternative models, i.e.,
dark energy/modified gravity models.
A guideline for modifying GR is given by the Lovelock's theorem~\cite{Lovelock:1971yv}.
The theorem states that the most general theory in four dimensions having at most
second-order Euler-Lagrange equations, respecting general covariance, and written
in terms of only the metric is nothing but GR with a cosmological constant.
Here, the second-order nature of field equations is desirable, since higher-order field equations generically lead to unstable extra degrees of freedom (DOFs) called Ostrogradsky ghosts~\cite{Woodard:2015zca,Motohashi:2020psc} unless the higher derivative terms are degenerate~\cite{Motohashi:2014opa,Langlois:2015cwa,Motohashi:2016ftl,Klein:2016aiq,Motohashi:2017eya,Motohashi:2018pxg}.\footnote{It is known that $f(R)$ gravity~\cite{Sotiriou:2008rp,DeFelice:2010aj} yields higher-order equations of motion.
Nevertheless, this theory is free of Ostrogradsky ghosts because it can be recast into 
GR with a canonical scalar field by field redefinition. 
}
Hence, a natural way to extend the framework of GR plus a cosmological constant
is to incorporate some new DOFs on top of the metric
in such a way that the action does not contain nondegenerate higher derivative interactions.
To capture aspects of such dark energy/modified gravity models,
it is useful to consider those having a single scalar field in addition to
the metric, i.e., the class of scalar-tensor theories.
Even in this restricted class, there exist innumerably many theories, so we need a comprehensive framework to treat them in a unified manner.
The Horndeski theory~\cite{Horndeski:1974wa,Deffayet:2011gz,Kobayashi:2011nu} is a well-known such comprehensive framework, since it is the most general scalar-tensor theory in four dimensions whose Euler-Lagrange equations are at most second-order.
By allowing for the existence of degenerate higher derivative terms in the Euler-Lagrange equations, the Horndeski theory is generalized to the Gleyzes-Langlois-Piazza-Vernizzi (GLPV, also known as beyond Horndeski) theory~\cite{Gleyzes:2014dya} and further to degenerate higher-order scalar-tensor (DHOST, also known as extended scalar-tensor) theories~\cite{Langlois:2015cwa,Crisostomi:2016czh,BenAchour:2016fzp,Takahashi:2017pje,Langlois:2018jdg}.
For recent reviews, see \cite{Langlois:2018dxi,Kobayashi:2019hrl}.

Generically, such healthy scalar-tensor theories possess three dynamical DOFs, two of which come from the metric and one from the scalar field.
However, there are some special models where only two DOFs are dynamical as in GR~\cite{Lin:2017oow,Chagoya:2018yna,Aoki:2018zcv,Afshordi:2006ad,Iyonaga:2018vnu,Gao:2019twq}.
In this sense, these models can be regarded as {\it minimal} modifications of GR, which could provide the second
most economical explanation of the accelerated expansion
next to the cosmological constant.
Indeed, the authors of \cite{Aoki:2018brq} showed that the model proposed in \cite{Aoki:2018zcv} can explain dark energy.
This model was obtained by performing a canonical transformation on GR, utilizing the idea that a canonical transformation preserves the number of physical DOFs~\cite{Domenech:2015tca,Takahashi:2017zgr}.
Along this line, we focus on the framework \cite{Iyonaga:2018vnu} of ourselves, which was invented as an extension of the cuscuton theory~\cite{Afshordi:2006ad}.
Regarding the original cuscuton model, its various aspects have been studied.
A Hamiltonian analysis was performed in \cite{Gomes:2017tzd}.
The cosmic microwave background and matter power spectra can be distinguished from those in GR~\cite{Afshordi:2007yx}.
Stable bounce cosmology~\cite{Boruah:2018pvq,Quintin:2019orx}, a reasonable power-law inflation model~\cite{Ito:2019fie}, and an accelerating universe with an extra dimension~\cite{Ito:2019ztb} based on the cuscuton have been studied.
It was also shown that the cuscuton theory with a quadratic potential can be considered as a low-energy limit of the (non-projectable) Ho\v{r}ava-Lifshitz theory~\cite{Afshordi:2009tt,Bhattacharyya:2016mah}.
The authors of \cite{deRham:2016ged} pointed out the absence of caustic singularities in cuscuton-like scalar field theories.
Moreover, the cuscuton admits extra symmetries other than the Poincar{\'e} symmetry~\cite{Pajer:2018egx,Grall:2019qof}.
These fascinating features of the cuscuton model motivated us to specify a broader class of scalar-tensor theories that inherit the two-DOF nature of the cuscuton, which we dubbed the ``extended cuscuton.''
The aim of this paper is to investigate
its cosmological aspects as to whether
the extended cuscuton can account for the current accelerated expansion
of the Universe.

The rest of this paper is organized as follows.
In \S \ref{sec_model}, we briefly explain the framework of extended cuscutons and present its action.
Then, in \S \ref{sec_cosmology}, we study cosmology in this class of models in the presence of a matter field.
We derive the background field equations and the quadratic action for scalar perturbations.
Also, we propose some requirements for the extended cuscutons to be a viable dark energy model.
In \S \ref{sec_quadraticcoupling}, we focus on an analytically solvable case and obtain criteria for the model to satisfy the viability requirements.
We find that this model can mimic the cosmological background evolution in the $\Lambda$CDM model, though the dynamics of the density fluctuations in general deviates from the one in the $\Lambda$CDM case.
Finally, we summarize our discussion in \S \ref{sec_conclusions}.

\section{The model}
\label{sec_model}

In \cite{Iyonaga:2018vnu}, the extended cuscuton model was obtained as a class of DHOST (more precisely, GLPV) theories in which the scalar field is nondynamical.
In the present paper, we focus on a subclass where the speed of gravitational waves, $c_{\rm GW}$, is equal to that of light, $c_{\rm light}(\coloneqq 1)$.
This is partly for simplicity and partly because the recent simultaneous observation of the gravitational waves GW170817 and the $\gamma$-ray burst 170817A emitted from a neutron star binary showed that $c_{\rm GW}$ coincides with $c_{\rm light}$ to a precision of $10^{-15}$ at least in the low-redshift universe~($z\lesssim 0.01$)~\cite{TheLIGOScientific:2017qsa,GBM:2017lvd,Monitor:2017mdv,Sakstein:2017xjx}.
This subclass is described by the following action:
    \begin{align}
    \label{eq_action}
    S_{\rm EC}=\int d^4x\sqrt{-g}\brb{G_2(\phi,X)+G_3(\phi,X)\Box\phi+G_4(\phi)R},
    \end{align}
with
\be
\begin{split}
	G_{2}
	&=u_{2}+v_{2} \sqrt{2 X}-\left(2 v_{3\phi}+4 v_{4\phi\phi}
		+\frac{3 v_{3}^{2}}{4 v_{4}}\right) X+\left(v_{3\phi}
		+2 v_{4\phi\phi}\right) X \log X, \\
	G_{3}&=-\left(\frac{v_{3}}{2}+v_{4\phi}\right) \log X, \quad
	G_{4}=v_{4},
\end{split}\label{action}
\ee
where $X\coloneqq -g^\mn\pa_\mu\phi\pa_\nu\phi/2$ and $R$ is the Ricci scalar.
Also, $u_{2}, v_{2}, v_{3}$, and $v_{4}$ are arbitrary functions of $\phi$, and a subscript~$\phi$ denotes a derivative with respect to $\phi$.
Note in passing that any model described by the action~\eqref{eq_action} satisfies $c_{\rm GW}=1$ around arbitrary backgrounds even without the ``cuscuton tuning''~\eqref{action} since it is conformally equivalent to general relativity with a scalar field in the form of kinetic gravity braiding~\cite{Deffayet:2010qz} (see also \cite{Kobayashi:2010cm}).
The appearance of $\log X$ in the action is one of the characteristic features of the extended cuscutons, which also appears in other contexts (see \cite{Pujolas:2011he,Afshordi:2014qaa} for examples).
Interestingly, our model is conformally equivalent to the one in \cite{Afshordi:2014qaa}.
Note also that the original cuscuton model proposed in \cite{Afshordi:2006ad} amounts to the choice~$v_3=0$ and $v_4=M_{\rm Pl}^2/2$.

We have three caveats on the physical DOFs of the extended cuscuton.
The first is about the relation between the DOFs and the homogeneity of the scalar field.
For the original cuscuton with timelike $\partial_\mu\phi$, the authors of \cite{Gomes:2017tzd} claimed that 
$\phi$ in general carries a scalar DOF and it vanishes only in the homogeneous limit.
However, this result is counterintuitive as one can always make the scalar field homogeneous, $\phi=\phi(t)$, by choosing the coordinate system appropriately (called the unitary gauge) when $\pa_\mu\phi$ is timelike.
We clarified this point in our previous paper~\cite{Iyonaga:2018vnu} by showing that the potentially existing scalar DOF actually does not propagate if an appropriate boundary condition is imposed.
Thus, provided that $\pa_\mu\phi$ is timelike, taking the unitary gauge does not change the number of physical DOFs, which allows us to choose this gauge in the following section.

The second is about the direction of $\partial_\mu\phi$.
The above action applies to situations with timelike $\partial_\mu\phi$ 
(i.e., $X>0$) so that $\sqrt{2X}$ and $\log X$ are real.
In order to incorporate cases with spacelike $\pa_\mu\phi$, one may replace $X\to|X|$.
In the resultant model, one finds that the number of dynamical DOFs depends on whether $\pa_\mu\phi$ is timelike or spacelike~\cite{Iyonaga:2018vnu}.
Specifically, when the gradient of the scalar field is spacelike, the scalar field remains dynamical as usual scalar-tensor theories.
This is similar to what happens in the 
spatially covariant gravity~\cite{Gao:2014soa} and U-degenerate theory \cite{DeFelice:2018mkq}, where a would-be unstable extra DOF becomes nondynamical when $\pa_\mu\phi$ is timelike.
As such, the scalar field breaks the Lorentz invariance and only the space diffeomorphisms remain.
There are many observational constraints on the Lorentz violation, e.g., from the Solar System tests~\cite{Blas:2010hb,Will:2014kxa} and more recently from binary black hole observations~\cite{Monitor:2017mdv,Gumrukcuoglu:2017ijh,Ramos:2018oku}.
These observational constraints should restrict our model, but it is beyond the scope of the present paper.

The third is about 
the existence of an extra half DOF.
As established in \cite{Iyonaga:2018vnu}, the model~\eqref{eq_action} is guaranteed to have less-than-three DOFs provided that the gradient of the scalar field is timelike.
The authors of \cite{Gao:2019twq} performed a more detailed Hamiltonian analysis to show that one needs an additional condition in general
to ensure the two-DOF nature, or otherwise there remains an extra half DOF.
The Ho\v{r}ava-Lifshitz gravity is one of the theories 
where the extra half DOF exhibits undesired behaviors~\cite{Blas:2009yd}.
For instance, the mode frequency of the half DOF diverges for static or spatially homogeneous backgrounds.
Also, the phase space of the Ho\v{r}ava-Lifshitz gravity is described by odd number of variables, which means that there is no symplectic structure.
The constraint structure of our extended cuscuton is similar to the Ho\v{r}ava-Lifshitz gravity, so something similar might happen to our model when there is an extra half DOF.
In order for the specific
model~\eqref{eq_action} to have exactly two DOFs, $v_4$ should be a nonvanishing constant.
However, we consider $\phi$-dependent $v_4$ in this paper,
as this potentially pathological
half DOF does not show up in the present cosmological setup.

\section{Cosmology}
\label{sec_cosmology}
\subsection{Background}
\label{subsec_background}
We study a homogeneous and isotropic universe in
the presence of a matter field, and hence consider the following action:
    \be
    S=S_{\rm EC}+\int d^4x\sqrt{-g}\mL_{\rm m}. \label{eq_action+m}
    \ee
The metric~$g_\mn$ and cuscuton~$\phi$ are assumed to have the form
\begin{align}
\label{eq_bguniverse}
    g_\mn dx^\mu dx^\nu=-N^{2}(t) d t^{2}+a^{2}(t) \delta_{i j} d x^{i} d x^{j}, \quad \phi=\phi(t).
\end{align}
In order to mimic barotropic perfect fluid, we write the matter Lagrangian~$\mL_{\rm m}$ in terms of a scalar field~$\chi$ as in \cite{ArmendarizPicon:1999rj},
\begin{align}
	\mL_{\rm m}=P(Y),\quad
	Y\coloneqq-\fr{1}{2}g^{\mu\nu}\pa_{\mu}\chi\pa_{\nu}\chi. \label{eq_Lm}
\end{align}
Here, $\chi$ is assumed to be a function of $t$ only.
Then, the energy density, pressure, and squared sound speed of $\chi$ are respectively written as
\begin{align}
	\rho_\mathrm{m}=2YP_{Y}-P,\quad
	p_\mathrm{m}=P,\quad
    c_{\rm s}^{2}=\fr{P_{Y}}{P_{Y} + 2YP_{YY}},
\end{align}
where $P_{Y}\coloneqq dP/dY$.
We substitute the ansatz~\eqref{eq_bguniverse} into the action~\eqref{eq_action+m}, from which we can derive the field equations for $N$, $a$, $\phi$, and $\chi$.
Among these EOMs, we focus on those for $N$, $a$, and $\phi$ since only three of the four EOMs are independent.
Note that the EOM for $N$ cannot be reproduced from the other EOMs.
Therefore, one may set $N=1$ only after deriving the EOM for $N$~\cite{Motohashi:2016prk}.
When we consider late-time cosmology where only the dust component is important, we may set $p_{\rm m}=0$ and $c_{\rm s}=0$.
One may naively think that this dust limit is ill-defined in the present case where we mimic perfect-fluid matter by \eqref{eq_Lm}, since $p_{\rm m}\to 0$ implies $\mL_{\rm m}=P(Y)\to 0$.
Nevertheless, once we rewrite every $P$ and its derivative in terms of $\rho_{\rm m}$, $p_{\rm m}$, and $c_{\rm s}$, we can safely take the dust limit~\cite{Boubekeur:2008kn}.

In deriving the field equations, we assume the time derivative of $\phi$ satisfies $\dot{\phi}>0$ to fix the branch of the square root originating from the term~$\sqrt{2X}$ in \eqref{action}.
One could in principle assume $\dot{\phi}<0$ instead, and in that case one should replace $v_2\to -v_2$ in the following analysis.
The Euler-Lagrange equations for $N$ and $a$ read, respectively,
    \begin{align}
    \mE_N&\coloneqq 6v_4H^2+u_2-3v_3H\dot{\phi}+\fr{3v_3^2}{8v_4}\dot{\phi}^2-\rho_{\rm m}=0, \label{eq_EOMN} \\
    \mE_a&\coloneqq 2v_4(3H^2+2\dot{H})+u_2+v_2\dot{\phi}+4v_{4\phi}H\dot{\phi}-\fr{3v_3^2}{8v_4}\dot{\phi}^2-v_{3\phi}\dot{\phi}^2-v_3\ddot{\phi}+p_{\rm m}=0. \label{eq_EOMa}
    \end{align}
The Euler-Lagrange equation for $\phi$ is written as
        \be
        \mE_\phi\coloneqq
        -\fr{3v_3^2}{4v_4}\ddot{\phi}+3v_3\dot{H}-\fr{9v_3^2}{4v_4}H\dot{\phi}
        -\fr{3v_3(2v_{3\phi}v_4-v_3v_{4\phi})}{8v_4^2}\dot{\phi}^2
        -u_{2\phi}+3v_2H+3H^2(3v_3+2v_{4\phi})=0. \label{eq_EOMphi}
        \ee
Taking a linear combination~$4v_4\mE_\phi-3v_3\mE_a$, one can simultaneously remove $\dot{H}$ and $\ddot{\phi}$ to obtain a constraint equation, which is a property of the extended cuscuton models.
Note that, when $v_3=0$, there is no $\dot{H}$ or $\ddot{\phi}$ in $\mE_\phi$ from the beginning.
It should also be noted that one can obtain the continuity equation~$\dot{\rho}_{\rm m}+3H(\rho_{\rm m}+p_{\rm m})=0$ by combining the EOMs~\eqref{eq_EOMN}, \eqref{eq_EOMa}, and \eqref{eq_EOMphi}.

In what follows, let us discuss some viability requirements for the present framework to serve as a dark energy model.
Later in \S \ref{sec_quadraticcoupling}, these requirements are used to constrain model parameters.
\begin{enumerate}
\renewcommand{\theenumi}{\Alph{enumi}}
\renewcommand{\labelenumi}{[\theenumi]}
    \item \label{reqA} \textit{Asymptotic behavior of the Hubble parameter}\\
    We require the following asymptotic behavior for the Hubble parameter:
    \begin{eqnarray}
    \label{H_constraints}
        \left\{
        \begin{array}{ll}
        H\to \mathrm{const}\cdot a^{-3/2}&\mathrm{for} ~t\to t_{\rm i}, \\
    	H\to \mathrm{const}\quad&\mathrm{for} ~t\to \infty,
        \end{array}
        \right.
    \end{eqnarray}
    so that it behaves as in the matter-dominated universe for $t\to t_{\rm i}$ (with $t_{\rm i}$ being some early initial time) and the de Sitter universe for $t\to \infty$.

    \item \label{reqB} \textit{Accelerating universe at the present time}\\
    Whether the universe is experiencing an accelerated expansion can be judged by
    looking at the Hubble slow-roll parameter~$\ep_H\coloneqq -\dot{H}/H^2$.
    Since $\ddot{a}\propto 1-\ep_H$, the accelerated (decelerated) expansion corresponds to $\ep_H<1$ ($\ep_H>1$).
    We require that the current value of $\ep_H$ should be less than unity.

    \item \label{reqC} \textit{Positive} $\dot{\phi}$\\
    Since we assumed $\dot{\phi}>0$ as mentioned above, we require that $\dot{\phi}$ must remain positive throughout its time evolution.

    \item \label{reqD} \textit{Positive nonminimal coupling function}\\
    A negative coupling to the Ricci scalar leads to unstable tensor perturbations.
    Moreover, it also results in negative Newton's constant, as we shall see in the next section.
    Therefore, we require that $G_{4}=2v_4(\phi)>0$.
\end{enumerate}

\subsection{Scalar perturbations}
\label{sec_perts}
To derive the evolution equation for the matter density fluctuations,
we consider scalar perturbations around the cosmological background \eqref{eq_bguniverse}.
We keep $p_{\rm m}$ and $c_{\rm s}$ for the moment,
and the dust limit is taken in the final step.
We write the metric as
    \be
    g_\mn dx^\mu dx^\nu
    =-(1+2\delta N)dt^2+2\pa_i\psi dtdx^i+a^{2}(1+2\zeta)\delta_{ij}dx^idx^j,
    \ee
where $\delta N$, $\psi$, and $\zeta$ are scalar perturbations.
Regarding the cuscuton field, as explained earlier, we can safely take the unitary gauge~$\phi=\phi(t)$.
The matter field also fluctuates as $\chi= \chi(t) + \del \chi(t,\vec{x})$, and $\del \chi$ is related to the gauge-invariant density fluctuation of the $\chi$ field as
	\be
	\delta=\fr{\rho_\mathrm{m}+p_\mathrm{m}}{\rho_\mathrm{m} c_{\rm s}^2}\bra{\fr{\dot{\del\chi}}{\dot{\chi}}-\delta N}+3\fr{\rho_\mathrm{m}+p_\mathrm{m}}{\rho_\mathrm{m}}\zeta. \label{delchi}
	\ee

Below, we work in the Fourier space.
To recast the real-space Lagrangian into the Fourier-space one, we first perform integration by parts so that each variable has an even number of spatial derivatives, followed by the replacement~$\pa^2\to -k^2$.
We then proceed to reexpress the Lagrangian in terms of $\del$ instead of $\del \chi$.
The Lagrangian contains the following terms associated with $\del\chi$:
	\be
	\mL\supset a^3\bra{\fr{\rho_\mathrm{m}+p_\mathrm{m}}{4c_{\rm s}^2Y}\dot{\del\chi}^2-\fr{\rho_\mathrm{m}+p_\mathrm{m}}{4Y}\fr{k^2}{a^2}\del\chi^2+\del\chi\cdot\xi},
	\ee
where $\xi$ denotes the terms that are linear in $\delta N$, $\psi$, and $\zeta$.
One can add the following term to $\mL$ without changing the dynamics:
	\be
	\mL_{\del\chi\to\del}=-a^3\fr{\rho_\mathrm{m}+p_\mathrm{m}}{4c_{\rm s}^2Y}\brc{\dot{\del\chi}-\dot{\chi}\brb{c_{\rm s}^2
  \bra{\fr{\rho_\mathrm{m}}{\rho_\mathrm{m}+p_\mathrm{m}}\del-3\zeta}+\delta N}}^2,
  \label{dtodchi}
	\ee
because upon substituting the solution to the Euler-Lagrange equation for $\delta$,
namely, Eq.~\eqref{delchi}, this Lagrangian vanishes.
Note that the overall normalization of~\eqref{dtodchi} is chosen so that
$\mL'\coloneqq \mL+\mL_{\del\chi\to\del}$ is linear in $\dot{\del\chi}$.
Consequently, one can eliminate $\del\chi$ by use of its EOM and we are left with the quadratic action written in terms of $(\delta N,\psi,\zeta,\del)$:
	\begin{align}
	\mL'=a^3&\biggl\{-6v_4\dot{\zeta}^2+\brb{2v_4\fr{k^2}{a^2}-\fr{9}{2}c_{\rm s}^2(\rho_\mathrm{m}+p_\mathrm{m})}\zeta^2-\fr{3\Theta^2}{2v_4}\delta N^2
	+2\Theta\fr{k^2}{a^2}\delta N\psi-4v_4\fr{k^2}{a^2}\psi\dot{\zeta}+6\Theta\delta N\dot{\zeta} \nonumber \\
	&~~~+\brb{4v_4\fr{k^2}{a^2}+3(\rho_\mathrm{m}+p_\mathrm{m})}\delta N\zeta+\fr{a^2\rho_\mathrm{m}^2}{2k^2(\rho_\mathrm{m}+p_\mathrm{m})}\brb{\dot{\del}+\fr{k^2(\rho_\mathrm{m}+p_\mathrm{m})}{a^2\rho_\mathrm{m}}\psi}^2 \nonumber \\
	&~~~-\fr{\rho_\mathrm{m}}{2(\rho_\mathrm{m}+p_\mathrm{m})}\bra{\rho_\mathrm{m} c_{\rm s}^2+\fr{3a^2}{k^2}\brc{5H^2(\rho_\mathrm{m} c_{\rm s}^2-p_\mathrm{m})+\fr{d}{dt}\brb{(\rho_\mathrm{m} c_{\rm s}^2-p_\mathrm{m})H}}}\del^2 \nonumber \\
	&~~~-\rho_\mathrm{m}\delta N\del+3H(\rho_\mathrm{m} c_{\rm s}^2-p_\mathrm{m})\psi\del+3\rho_\mathrm{m} c_{\rm s}^2\zeta\del\biggr\},
	\end{align}
where we have defined
    \be
    \Theta\coloneqq 2v_4H-\fr{1}{2}v_3\dot{\phi}.
    \ee
Eliminating $\delta N$ and $\psi$ by the
use of their EOMs, the Lagrangian can be written in the form
	\be
	\mL''=a^3\brb{a_1(t,k)\dot{\del}^2+a_2(t,k)\del^2+2a_3(t)\zeta\del+a_4(t,k)\zeta^2}.
	\ee
Finally, by integrating out $\zeta$, we obtain the quadratic action for $\delta$ as
	\be
	\mL_\delta=a^3\bra{\mA\dot{\del}^2+\mB\del^2},
	\ee
from which we obtain the evolution equation for $\delta$ as follows:
    \be
    \ddot{\delta}+\bra{3H+\fr{\dot{\mA}}{\mA}}\dot{\delta}-\fr{\mB}{\mA}\delta=0. \label{eq_EOMdelta}
    \ee
A caveat should be added here.
In the case of generic scalar-tensor theories where the scalar field is dynamical, we still have an additional dynamical DOF other than $\delta$ at this stage.
In order to extract the effective dynamics of the density fluctuations on subhorizon scales, one usually makes the quasi-static approximation.
In the present case of the extended cuscutons, however, the quadratic action is written solely in terms of the density fluctuations even before taking the subhorizon limit.
This is one of the distinct properties of cuscuton-like theories.

In what follows, we consider a dust fluid 
by taking the limits~$p_{\rm m}\to 0$ and $c_{\rm s}\to 0$.\footnote{This limiting procedure is justified in \cite{DeFelice:2015moy,Babichev:2018twg}.
Instead, one may consider the action for a dust fluid from the beginning~\cite{Brown:1994py}.}
Then, the coefficients~$\mA$ and $\mB$ are respectively written as
    \be
    \mA=\fr{2v_4\rho_{\rm m}}{4v_4+3(a^2/k^2)\rho_{\rm m}}\fr{a^2}{k^2}, \quad
    \mB=
	\fr{2\lambda v_4\rho_{\rm m}^2}{
  [4v_4+3(a^2/k^2)\rho_\mathrm{m}]^2}
  \fr{a^2}{k^2},
    \ee
with
    \be
    \la\coloneqq \fr{4v_4\brb{2\bra{\dot{v}_4+v_4H}^2-v_4\bra{\rho_\mathrm{m}+2\dot{\Theta}+2H\Theta}}
    -3(a^2/k^2)\rho_\mathrm{m}\brb{v_4
    \bra{\rho_\mathrm{m}+2\dot{\Theta}}-2\dot{v}_4\Theta}}
	{4v_4\brb{\Theta\bra{4\dot{v}_4-\Theta}-v_4\bra{\rho_{\rm m}+2\dot{\Theta}-2H\Theta}}
    -3(a^2/k^2)\rho_\mathrm{m}\brb{v_4
    \bra{\rho_\mathrm{m}+2\dot{\Theta}}-2\dot{v}_4\Theta}}.
    \ee
In the subhorizon limit, Eq.~\eqref{eq_EOMdelta} reduces to the following form:
    \be
    \ddot{\delta}+2H\dot \delta-4\pi G_{\rm eff}\rho_{\rm m}\delta=0,
    \ee
where we have defined the effective gravitational coupling~$G_{\rm eff}$
for the density fluctuations as
    \be
    4\pi G_{\rm eff}\coloneqq \lim_{k\to\infty}\fr{\mB}{\rho_{\rm m}\mA}
    =\fr{1}{4v_4}
    \brb{1+\fr{\bra{2\dot{v}_4+2v_4H-\Theta}^2}{\Theta\bra{4\dot{v}_4-\Theta}-v_4\bra{\rho_{\rm m}+2\dot{\Theta}-2H\Theta}}}. \label{eq_Geff}
    \ee
The Poisson equations for the gauge-invariant gravitational
potentials, $\Psi=\delta N+\dot{\psi}$ and $\Phi=-\zeta-H\psi$, are
given by
    \be
    -\fr{k^2}{a^2}\Psi=4\pi G_{\rm eff}\rho_{\rm m}\delta, \quad
    -\fr{k^2}{a^2}\Phi=4\pi \bar{G}_{\rm eff}\rho_{\rm m}\delta,
    \ee
where $\bar{G}_{\rm eff}$ is defined by
    \be
    4\pi \bar{G}_{\rm eff}\coloneqq \fr{1}{4v_4}
    \brb{1+\fr{\bra{2v_4H-\Theta}\bra{2\dot{v}_4+2v_4H-\Theta}}{\Theta\bra{4\dot{v}_4-\Theta}-v_4\bra{\rho_{\rm m}+2\dot{\Theta}-2H\Theta}}}. \label{eq_Geffb}
    \ee
Note that, if and only if $\dot v_4(2\dot{v}_4+2v_4H-\Theta)=0$,
i.e., $v_{4\phi}=0$ or $v_3+4v_{4\phi}=0$, we have
$G_{\rm eff}=\bar{G}_{\rm eff}$ so that
the so-called
gravitational slip parameter~$\eta\coloneqq \Psi/\Phi$ is equal to unity as in GR.\footnote{
As was shown in \cite{Saltas:2014dha}, the deviation of the slip parameter from unity is characterized by the functions called $\alpha_\mathrm{M}$ and $\alpha_\mathrm{T}$, which are fixed once the arbitrary functions in the action~\eqref{eq_action} are fixed.
Specifically, the slip parameter becomes unity if and only if $\alpha_\mathrm{M}=\alpha_\mathrm{T}=0$.
On the other hand, for our model satisfying \eqref{action}, we have $\alpha_\mathrm{T}=0$ and $\alpha_\mathrm{M}\propto \dot{G}_4=v_{4\phi}\dot{\phi}\ne 0$ in general, and thus the slip parameter deviates from unity.
Therefore, our result is consistent with the one in \cite{Saltas:2014dha}.
}

It is important to see the difference between the above effective gravitational
coupling for linear density fluctuations and the locally measured value of
Newton's constant, $G_N$. To evaluate $G_N$ in the extended cuscuton
theory, one can closely follow the discussion for the Vainshtein solution
of~\cite{Kimura:2011dc}. Although $\phi$ is not dynamical in the
present setup due to the particular choice of the functions in the action~\eqref{eq_action}, this ``cuscuton tuning''
does not change the procedure to derive a static and spherically
symmetric solution in the weak gravity regime.
Thus, regardless of whether $\phi$ is dynamical or not, its nonlinearities
play an essential role below a certain scale to reproduce Newtonian gravity,
provided that $G_{3X}\neq 0$.
It then follows that $G_N$ is given by~\cite{Kimura:2011dc}\footnote{Some
assumptions on the size of various coefficients are made in~\cite{Kimura:2011dc}.
All these assumptions are valid as well in the extended cuscuton theory
if it accounts for the present accelerated expansion of the Universe.}
    \be
    4\pi G_N=\fr{1}{4v_4}, \label{eq_GN}
    \ee
which is different from $G_{\rm eff}$ as long as $v_3+4v_{4\phi}\ne 0$.
Note that $G_N$ depends on time and is not actually a constant since $v_4$ is a function of $\phi$, which varies in time.

To sum up, although $\phi$ is nondynamical in the extended cuscuton theory,
the evolution of density fluctuations is modified in the same way as
in usual scalar-tensor theories.

\section{Exactly solvable model}
\label{sec_quadraticcoupling}
In the previous section, we obtained the background field equations, the effective gravitational coupling~$G_\mathrm{eff}$, and the Newton's constant~$G_N$ for generic models described by \eqref{eq_action}.
Now, we turn to more specific discussions using a simple subclass which can be solved analytically.

\subsection{The Lagrangian and basic equations}
\label{subsec_equations}
We consider the extended cuscuton theory with a quadratic nonminimal coupling,
    \be
    S_{\rm EC}=\int d^4x\sqrt{-g}
    \brb{\bra{\frac{M_*^2}{2}+\mu \phi^2}R
    -\frac{1}{2}m^{2}\phi^{2}+(\alpha+\beta\phi)\sqrt{2X}
    +4\mu X(-2+\log X)-2\mu\phi \log X \Box \phi}, \label{eq_model}
    \ee
which corresponds to the following choice of the functions in \eqref{action}:
\begin{align}
\label{quadratic_model}
	u_{2}=-\frac{1}{2}m^{2}\phi^{2},\quad
	v_{2}=\alpha+\beta\phi,\quad
	v_{3}=0,\quad
	v_{4}=\frac{M_*^2}{2}+\mu \phi^2.
\end{align}
Here, $M_*$, $\mu$, $\alpha$, $\beta$, and $m$ are nonvanishing constant.
Note that the original cuscuton corresponds to the limit~$\mu\to 0$ and $\beta\to 0$, and hence the terms with $\mu$ or $\beta$ characterize the difference from the original model.
Note that nonvanishing $\mu$ leads to $G_{3X}\neq 0$, meaning that Newtonian gravity is reproduced except for the time dependence of $G_N$.
The field equations read
\begin{align}
\label{Friedmann2}
  \mathcal{E}_N
	&= 3(M_*^2+2\mu\phi^2)H^2 -\frac{1}{2}m^2 \phi^{2}-\rho_\mathrm{m}
	=0,\\
\label{wrt_a2}
	\mathcal{E}_a
	&= (M_*^2+2\mu\phi^2)(3H^2 + 2\dot{H}) -\frac{1}{2}m^2\phi^2 + (\alpha+\beta\phi)\dot{\phi} + 8\mu H\phi\dot{\phi}
	=0,\\
\label{wrt_phi2}
 	\mathcal{E}_\phi
	&= 3(\alpha+\beta\phi) H + (m^{2}+12\mu H^{2})\phi
	=0,
\end{align}
where we have set $p_{\rm m}=0$.
We use the redshift~$z\coloneqq a(t_0)/a(t)-1$ (with $t_0$ being the present time) as the time coordinate.
Provided that the scale factor is monotonically increasing from zero to infinity in time, then $z=\infty$ corresponds to the initial time and $z=-1$ formally corresponds to the infinite future.
Let us define the following dimensionless variables:
\begin{align}
	M\coloneqq \frac{H_0^2}{m^2}\mu,\quad
	A\coloneqq \frac{\alpha}{m M_*},\quad
	B\coloneqq \frac{H_0}{m^2}\beta,\quad
	\hat{\phi}(z)\coloneqq \frac{m}{M_* H_0}\phi(z),\quad
	\hat{H}(z)\coloneqq \frac{H(z)}{H_0}, \label{eq_paramND}
\end{align}
where $H_0 \coloneqq H(z=0)$.
In terms of the dimensionless variables, Eqs.~\eqref{wrt_a2} and \eqref{wrt_phi2} are rewritten as
    \begin{align}
    \fr{\mE_a}{M_*^2H_0^2}&=(1+2M\hat{\phi}^2)\hat{H}\brb{3\hat{H}-2(1+z)\hat{H}'}-\fr{1}{2}\hat{\phi}^2-(A+B\hat{\phi}+8M\hat{H}\hat{\phi})(1+z)\hat{H}\hat{\phi}'=0, \label{EOMaND} \\
    \fr{\mE_\phi}{mM_* H_0}&=3A\hat{H}+(1+3B\hat{H}+12M\hat{H}^2)\hat{\phi}=0, \label{EOMphiND}
    \end{align}
where a prime denotes a derivative with respect to $z$.
Removing $\hat{\phi}$ from \eqref{EOMaND} by using
\eqref{EOMphiND}, we are left with the following first-order differential equation for $\hat{H}$:
    \begin{align}
    (1+z)\hat{H}'
    =\fr{3\hat{H}}{2}
    \fr{(1+3B\hat{H}+12M\hat{H}^2)\brb{2(1+3B\hat{H}+12M\hat{H}^2)^2-3A^2(1-12M\hat{H}^2)}}
    {2(1+3B\hat{H}+12M\hat{H}^2)^3-3A^2(1-36M\hat{H}^2-36MB\hat{H}^3)}. \label{eq_H}
    \end{align}
Note in passing that $1+3B\hat{H}+12M\hat{H}^2\ne 0$ should be required for any $z$ so that \eqref{EOMphiND} can always be solved for $\hat{\phi}$.
Note also that, in the limit~$\hat{H}\to \infty$, Eq.~\eqref{eq_H} takes the form
    \be
    (1+z)\hat{H}'=\fr{3\hat{H}}{2}, \label{eq_HinMD}
    \ee
which yields the desired behavior of the Hubble parameter at early times, namely, $H\to {\rm const}\cdot a^{-3/2}\propto (1+z)^{3/2}$.

Equation~\eqref{Friedmann2} is used to determine
the matter energy density $\rho_{\rm m}$. In terms of the matter density parameter~$\Omega_{\rm m0}\coloneqq 8\pi G_N\rho_{\rm m}/3H^2|_{z=0}$, Eq.~\eqref{Friedmann2}
can be written as
\begin{align}
\Omega_{\rm m0}=1-\frac{3A^2}{2[(1+3B+12M)^2+18MA^2]},\label{omegam0}
\end{align}
showing that $\Omega_{\rm m0}$ is fixed by the parameters $M$, $A$, and $B$.

We will see that \eqref{eq_H} can be solved analytically.
However, before proceeding let us look for the parameter region
that fulfills the requirements~[\ref{reqA}]--[\ref{reqD}]
in order for \eqref{eq_model} to be a viable dark energy model.

\subsection{Viable parameter region}
\label{subsec_viability}

Now we apply the requirements~[\ref{reqA}]--[\ref{reqD}] mentioned in \S \ref{subsec_background} to the present case and find the viable region in the three-dimensional parameter space~$(M,A,B)$ by studying the dynamics of $\hat{H}$ based on \eqref{eq_H}.

We first demand [\ref{reqA}], namely, we require that $\hat{H}$ starts from a large value at some early initial time and approaches to a constant (which we denote by $\hat{H}_{\rm dS}$) in the infinite future.
Then, the asymptotic value~$\hat{H}_{\rm dS}$ should correspond to the largest stable equilibrium point of \eqref{eq_H}.\footnote{Here, an equilibrium point~$\hat{H}=\hat{H}_\ast$ is said to be stable if and only if $\hat{H}'<0$ (i.e., $d\hat{H}/dt>0$) for $\hat{H}\in (\hat{H}_\ast-\ep,\hat{H}_\ast)$ and $\hat{H}'>0$ (i.e., $d\hat{H}/dt<0$) for $\hat{H}\in (\hat{H}_\ast,\hat{H}_\ast+\ep)$, with $\ep$ being an infinitesimal positive number.}
Given that $\hat{H}>0$ and $1+3B\hat{H}+12M\hat{H}^2\ne 0$, $\hat{H}_{\rm dS}$ is given by one of the positive solutions (if they exist) of the following quartic equation:
    \be
    2(1+3B\hat{H}+12M\hat{H}^2)^2-3A^2(1-12M\hat{H}^2)=0. \label{eq_HdS}
    \ee
Provided that this equation has positive solutions, the largest one is a candidate of $\hat{H}_{\rm dS}$.

Let us now demand [\ref{reqC}], which is equivalent to $\hat{\phi}'<0$ since $\hat{\phi}'\propto -\dot{\phi}$.
Using \eqref{EOMphiND}, $\hat{\phi}'$ is written as
    \be
    \hat{\phi}'=-\fr{3A(1-12M\hat{H}^2)\hat{H}'}{(1+3B\hat{H}+12M\hat{H}^2)^2}.
    \ee
When $M$ is positive, the factor~$1-12M\hat{H}^2$ should be negative definite as otherwise $\hat{\phi}'$ changes sign during its evolution.
However, this contradicts the fact that $\hat{H}$ travels to $\hat{H}_{\rm dS}$ because
    \be
    1-12M\hat{H}_{\rm dS}^2=\fr{2(1+3B\hat{H}_{\rm dS}+12M\hat{H}_{\rm dS}^2)^2}{3A^2}>0.
    \ee
Hence, in what follows, we require $M<0$.
In this case, one can show that \eqref{eq_HdS} has at least one positive solution and that the largest solution provides a stable equilibrium point of \eqref{eq_H}.
Then, this largest solution can be identified as $\hat{H}_{\rm dS}$.
One can also verify that $\hat{H}'>0$ for $\hat{H}>\hat{H}_{\rm dS}$, and therefore one always has $\hat{\phi}'<0$ as long as $A>0$.
Moreover, we require $\hat{H}_{\rm dS}<1$ so that the evolution of $\hat{H}$ is consistent with the condition~$\hat{H}(z=0)=1$.
Given that $M<0$, the requirement~$\hat{H}_{\rm dS}<1$ is satisfied if
    \be
    1+3B+12M<0,\quad
    \fr{2(1+3B+12M)^2}{3A^2(1-12M)}>1.
    \ee

Regarding [\ref{reqD}], it is trivially satisfied as
    \be
    \fr{2v_4}{M_*^2}=1+\fr{18MA^2\hat{H}^2}{(1+3B\hat{H}+12M\hat{H}^2)^2}
    >1+\fr{18MA^2\hat{H}_{\rm dS}^2}{(1+3B\hat{H}_{\rm dS}+12M\hat{H}_{\rm dS}^2)^2}
    =\fr{1}{1-12M\hat{H}_{\rm dS}^2}>0.
    \ee
Thus, the requirement~[\ref{reqD}] does not narrows down the viable parameter region.

Finally, let us consider [\ref{reqB}].
The present value of the Hubble slow-roll parameter is written as
    \be
    \ep_H(z=0)=\hat{H}'(z=0)
    =\fr{3(1+3B+12M)\brb{2(1+3B+12M)^2-3A^2(1-12M)}}
    {2\brb{2(1+3B+12M)^3-3A^2(1-36M-36MB)}}.
    \ee
Requiring $\ep_H(z=0)<1$ to guarantee the accelerated expansion of the Universe at the present time, we have
    \be
    \fr{3A^2\brb{1+72M-432M^2+9B(1-4M)}}{2(1+3B+12M)^3}>1.
    \ee

\begin{figure}[h]
  \begin{minipage}[b]{0.45\linewidth}
    \centering
    \includegraphics[clip,width=7.0cm]{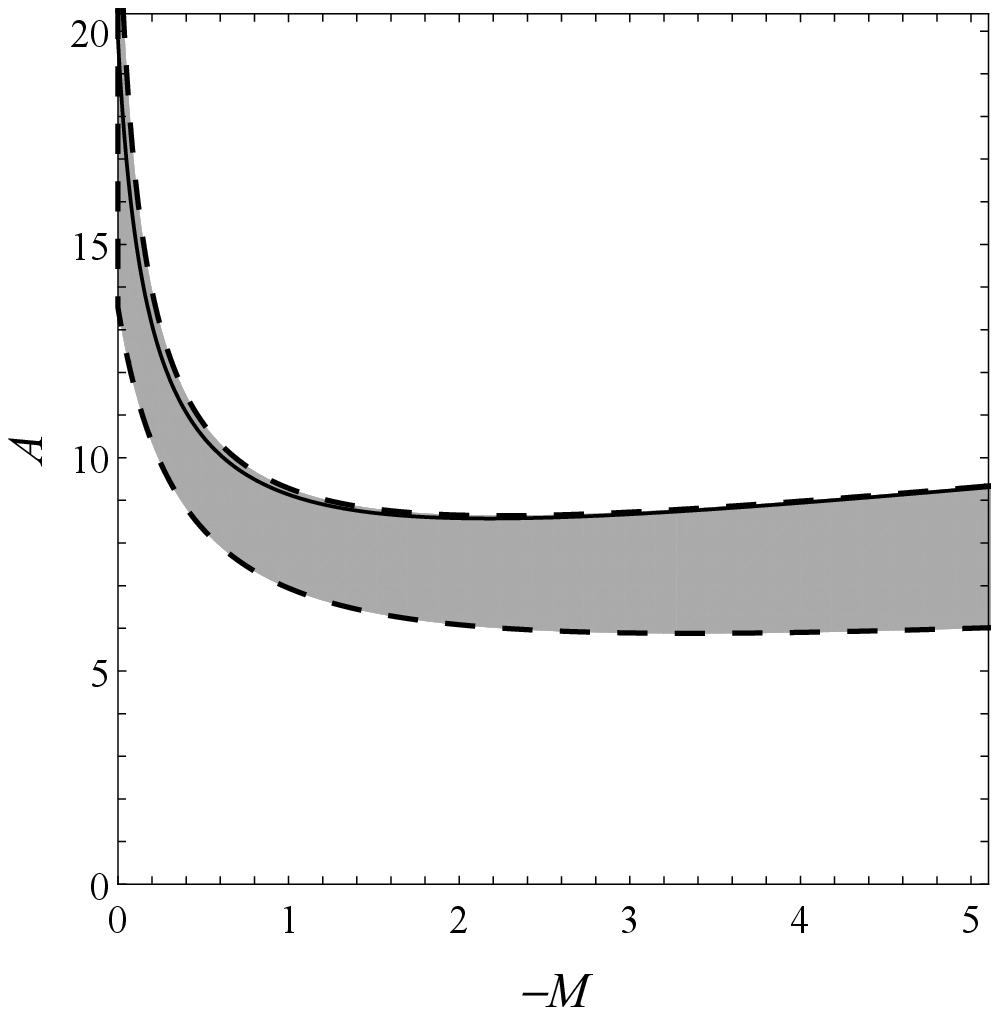}
    \subcaption{$B=-10$}\label{fig_viable_B=-10}
  \end{minipage}
  \begin{minipage}[b]{0.45\linewidth}
    \centering
    \includegraphics[clip,width=7.0cm]{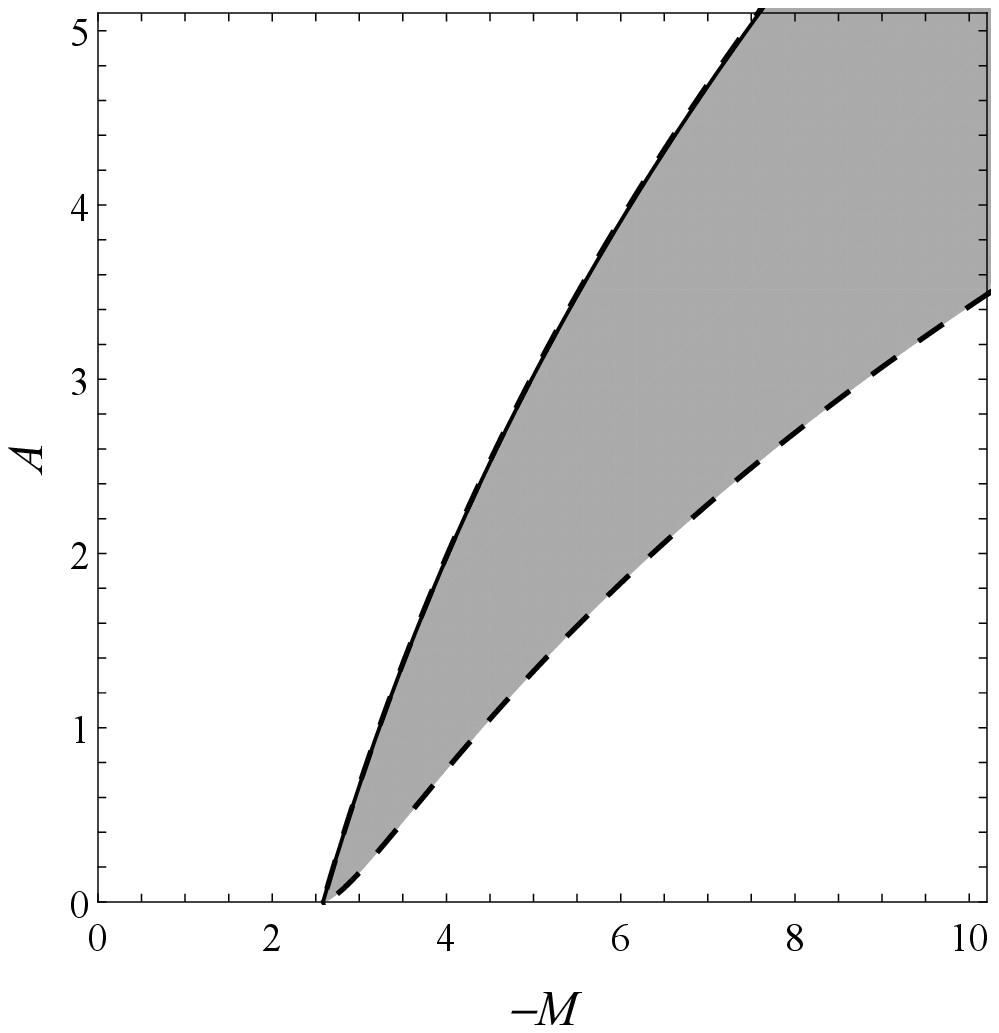}
    \subcaption{$B=10$}\label{fig_viable_B=10}
  \end{minipage}
\captionsetup{justification=RaggedRight}
  \caption{
  Two-dimensional sections of the parameter space~$(M,A,B)$ satisfying \eqref{eq_viabilityC} are colored gray (the boundary is indicated by dashed curves).
  The solid curves correspond to the parameters that yield $\Omega_{\rm m0}=0.3$, which almost overlap with the upper dashed curves.
  }
  \label{fig_viable}
\end{figure}

In summary, the requirements~[\ref{reqA}]--[\ref{reqD}] are satisfied if the following four conditions are fulfilled:
    \be
    M<\min \bra{0,-\fr{1+3B}{12}},\quad
    A>0,\quad
    \fr{2(1+3B+12M)^2}{3A^2(1-12M)}>1, \quad
    \fr{3A^2\brb{1+72M-432M^2+9B(1-4M)}}{2(1+3B+12M)^3}>1. \label{eq_viabilityC}
    \ee
We present two-dimensional sections of the viable parameter region at some fixed values of $B$ in Fig.~\ref{fig_viable}.

The matter density parameter~$\Omega_{\rm m0}$ is given in terms of $M$, $A$, and $B$ as \eqref{omegam0}.
For a fiducial value $\Omega_{\rm m0}=0.3$, Eq.~\eqref{omegam0} defines a two-dimensional surface in the parameter space~$(M,A,B)$, which appears as the solid curves in Fig.~\ref{fig_viable}.
For the parameters in the vicinity of these curves, one expects to have a background cosmological evolution that is similar to the one in the currently viable $\Lambda$CDM model.

\subsection{The solution}
\label{subsec_sol}

Having obtained the viable parameter region,
now we are in a position to analyze the exact solution to \eqref{eq_H}.
It is straightforward to integrate \eqref{eq_H}
to obtain the following algebraic equation for $\hat{H}$:
    \be
    \hat{H}^2\brb{2(1+3B\hat{H}+12M\hat{H}^2)^2-3A^2(1-12M\hat{H}^2)}+C(1+z)^3(1+3B\hat{H}+12M\hat{H}^2)^2=0, \label{eq_analytical}
    \ee
where the integration constant~$C$ is determined from $\hat{H}(z=0)=1$ as
    \be
    C=-2+\fr{3A^2(1-12M)}{(1+3B+12M)^2}. \label{eq_C}
    \ee
Note that \eqref{eq_HdS} is recovered in the limit $z\to -1$.

The Newton's constant~\eqref{eq_GN} and the effective gravitational coupling~\eqref{eq_Geff} are given, respectively, by
\begin{align}
  8\pi G_NM_*^2
  &=1-\fr{18MA^2\hat{H}^2}{(1+3B\hat{H}+12M\hat{H}^2)^2+18MA^2\hat{H}^2}, \label{eq_GNy} \\
  8\pi G_{\rm eff}M_*^2
  &=8\pi G_NM_*^2+\fr{864M^2A^2(1-12M\hat{H}^2)(1+3B\hat{H}+12M\hat{H}^2)\hat{H}^3}{(1-36M\hat{H}^2)\brb{(1+3B\hat{H}+12M\hat{H}^2)^2+18MA^2\hat{H}^2}^2}(1+z)\hat{H}', \label{eq_Geffy}
\end{align}
One can draw some information on the asymptotic behavior of
these quantities from~\eqref{eq_Geffy} and~\eqref{eq_GNy}.
In the infinite future, we have $\hat{H}'\to 0$, and thus $G_{\rm eff}/G_N\to 1$,
while for large $z$ where $\hat{H}\propto (1+z)^{3/2}$, we have
    \be
    8\pi G_{\rm eff}M_*^2\to 1+\fr{A^2}{8M\hat{H}^2}, \quad
    8\pi G_NM_*^2\to 1-\fr{A^2}{8M\hat{H}^2}.
    \ee

\begin{figure}[h]
 \begin{center}
    \includegraphics[clip,width=7.0cm]{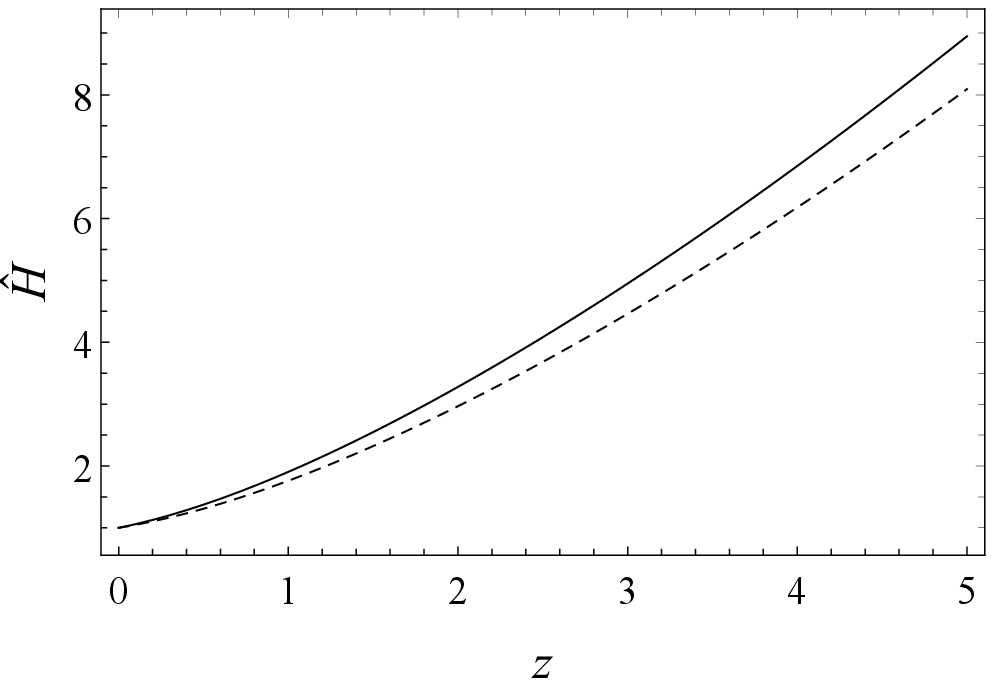}
    \hspace{10mm}
    \includegraphics[clip,width=7.0cm]{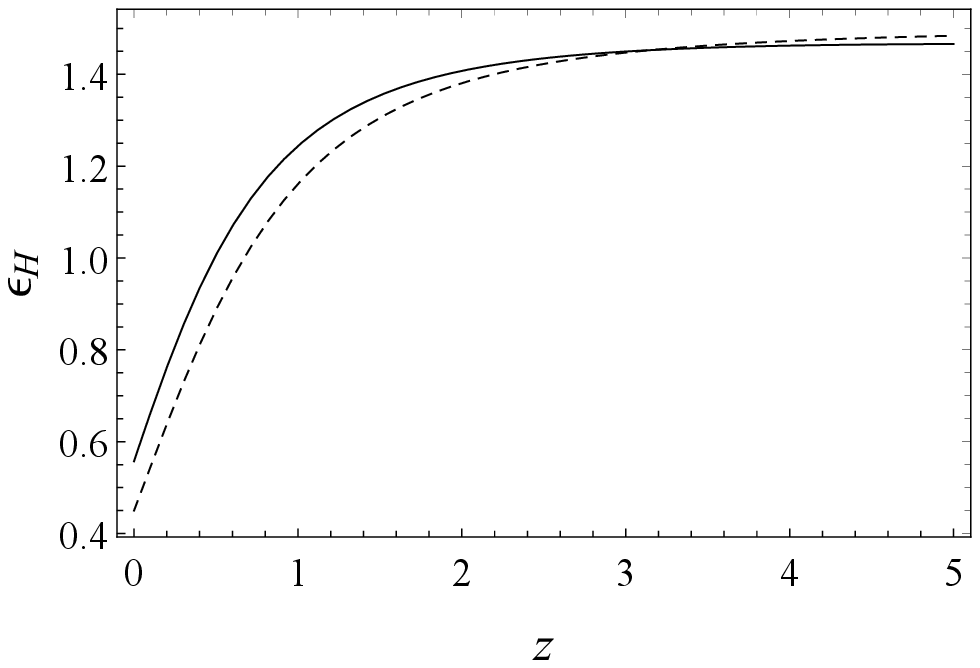}\\
    \vspace{5mm}
    \includegraphics[clip,width=7.0cm]{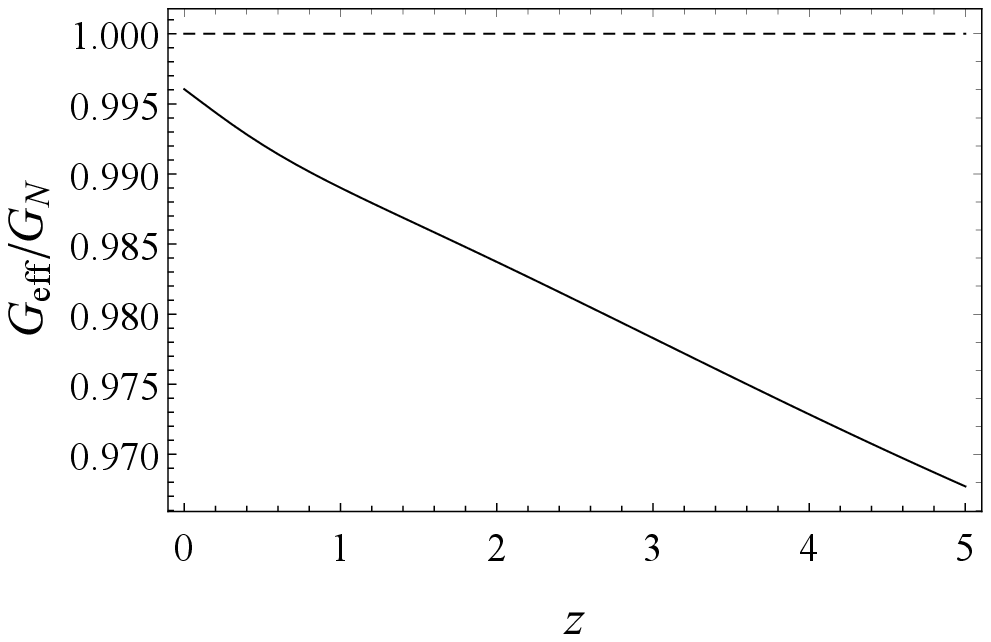}
    \hspace{10mm}
    \includegraphics[clip,width=7.0cm]{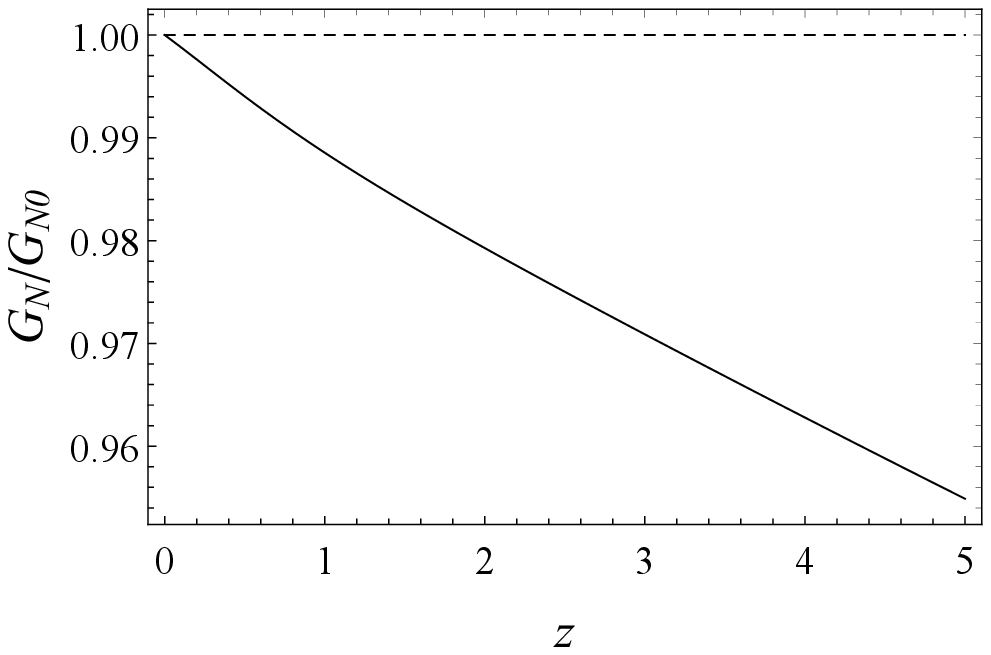}
\captionsetup{justification=RaggedRight}
    \caption{
    Time evolution of $\hat{H}$, $\epsilon_H$, $G_{\rm eff}/G_N$, and $G_N/G_{N0}$, with $G_{N0}\coloneqq G_N(z=0)$.
    The solid lines correspond to $(M,A,B)=(-0.03,17,-10)$ and the dashed lines represent the result of the $\Lambda$CDM model with $\Omega_{\rm m0}=0.3$.
    }
     \label{fig_G_wDE}
  \end{center}
\end{figure}

As an illustrative example, we plot the evolution of $\hat H$, $\epsilon_H$, and the gravitational couplings for $(M,A,B)=(-0.03,17,-10)$ in Fig.~\ref{fig_G_wDE}.
Note that this parameter choice fulfills the viability conditions~\eqref{eq_viabilityC} (see Fig.~\ref{fig_viable_B=-10}).
From these examples, we see that the background evolution is similar to the conventional $\Lambda$CDM model, while the evolution of the density fluctuations can be used to test the extended cuscuton as dark energy.
The time variation of Newton's constant can also be used to constrain the model, which, in the present case, is given by
\begin{align}
\left.\frac{\dot G_N}{HG_N}\right|_{z=0}
=\frac{54MA^2(1-12M)\brb{2(1+3B+12M)^2-3A^2(1-12M)}}
{\brb{(1+3B+12M)^2+18MA^2}\brb{2(1+3B+12M)^3-3A^2(1-36M-36MB)}},
\end{align}
while the observational bound reads $|\dot G_N/G_N|<0.02 H_0$~\cite{Williams:2004qba}.
One can check that the parameter choice~$(M,A,B)=(-0.03,17,-10)$ satisfies this bound.

Before proceeding to the concluding section, let us mention some limiting cases where one of the model parameters in \eqref{eq_model} is vanishing.
When $\alpha=0$ (i.e., $A=0$), we obtain $\phi=0$ from \eqref{wrt_phi2}, which contradicts the assumption that $\pa_\mu\phi$ is timelike (see \S \ref{sec_model}).
On the other hand, when $\mu=0$ (i.e., $M=0$), we obtain $G_N=G_{\rm eff}=(8\pi M_*^2)^{-1}$ from \eqref{eq_GNy} and \eqref{eq_Geffy}, while the spacetime and the cuscuton field can evolve in a nontrivial manner.

\section{Conclusions}
\label{sec_conclusions}
The extended cuscuton model is a general class of DHOST theories having a nondynamical scalar field when $\pa_\mu\phi$ is timelike.
In \S \ref{sec_cosmology}, we studied homogeneous and isotropic cosmology in the extended cuscutons described by the action~\eqref{eq_action} in the presence of a matter field.
We derived the background field equations and proposed the requirements~[\ref{reqA}]--[\ref{reqD}] for these theories to serve as a viable dark energy model.
Also, we investigated scalar perturbations to derive the evolution equation for the density fluctuations and the gravitational Poisson equations.
In \S \ref{sec_quadraticcoupling}, we turned to more specific discussions using a simple model~\eqref{eq_model} that can be solved analytically.
The model parameter~$\alpha$ appears as a coefficient of $\sqrt{2X}$, which is typical in the original cuscuton model.
On the other hand, the parameters~$\mu$ and $\beta$ characterize the difference from the original model.
In order to avoid technical complexity, we defined dimensionless parameters~$M$, $A$, and $B$, corresponding to $\mu$, $\alpha$, and $\beta$, respectively.
We obtained the viable region in the parameter space~$(M,A,B)$ which satisfies the requirements~[\ref{reqA}]--[\ref{reqD}].
We also plotted the evolution of the dimensionless Hubble parameter~$\hat{H}$, the Hubble slow-roll parameter~$\epsilon_H$, the ratio of the effective gravitational coupling~$G_\mathrm{eff}$ to the Newton's constant~$G_N$, and $G_N$ normalized by its present value for the parameter choice~$(M,A,B)=(-0.03,17,-10)$, which lies in the viable parameter region.
We found that the background evolution in this model can mimic the conventional $\Lambda$CDM model while the evolution of the density fluctuation deviates from the one in the $\Lambda$CDM case.
Moreover, this set of parameters satisfies the observational constraint on the time variation of the Newton's constant, $|\dot{G}_N/G_N|<0.02 H_0$.
Hence, one can test the extended cuscuton as dark energy by observations associated with the density fluctuations, e.g., the integrated Sachs-Wolfe effect or weak gravitational lensing, which we leave for future study.

As mentioned in \S \ref{sec_model}, in general, extended cuscutons have an extra half DOF on top of two tensor modes~\cite{Gao:2019twq}.
Nevertheless, at least up to linear perturbations on a homogeneous and isotropic background, we found no pathology caused by this half DOF.
However, we may encounter some inconsistencies in higher-order perturbations or on another background.
We hope to discuss this point in the near future.

\acknowledgements{
We would like to thank Zhi-Bang Yao for fruitful discussions.
This work was supported in part by
the Rikkyo University Special Fund for Research (A.I.),
JSPS KAKENHI Grant Nos.\ JP17H02894 and JP17K18778 (K.T.),
JSPS Bilateral Joint Research Projects (JSPS-NRF Collaboration) ``String Axion Cosmology'' (K.T.),
MEXT KAKENHI Grant Nos.\ JP17H06359, JP16K17707, and JP18H04355 (T.K.).
}

\bibliographystyle{mybibstyle}
\bibliography{cuscutoncos}

\begin{thebibliography}{67}%
\makeatletter
\providecommand \@ifxundefined [1]{%
 \@ifx{#1\undefined}
}%
\providecommand \@ifnum [1]{%
 \ifnum #1\expandafter \@firstoftwo
 \else \expandafter \@secondoftwo
 \fi
}%
\providecommand \@ifx [1]{%
 \ifx #1\expandafter \@firstoftwo
 \else \expandafter \@secondoftwo
 \fi
}%
\providecommand \natexlab [1]{#1}%
\providecommand \enquote  [1]{``#1''}%
\providecommand \bibnamefont  [1]{#1}%
\providecommand \bibfnamefont [1]{#1}%
\providecommand \citenamefont [1]{#1}%
\providecommand \href@noop [0]{\@secondoftwo}%
\providecommand \href [0]{\begingroup \@sanitize@url \@href}%
\providecommand \@href[1]{\@@startlink{#1}\@@href}%
\providecommand \@@href[1]{\endgroup#1\@@endlink}%
\providecommand \@sanitize@url [0]{\catcode `\\12\catcode `\$12\catcode
  `\&12\catcode `\#12\catcode `\^12\catcode `\_12\catcode `\%12\relax}%
\providecommand \@@startlink[1]{}%
\providecommand \@@endlink[0]{}%
\providecommand \url  [0]{\begingroup\@sanitize@url \@url }%
\providecommand \@url [1]{\endgroup\@href {#1}{\urlprefix }}%
\providecommand \urlprefix  [0]{URL }%
\providecommand \Eprint [0]{\href }%
\providecommand \doibase [0]{http://dx.doi.org/}%
\providecommand \selectlanguage [0]{\@gobble}%
\providecommand \bibinfo  [0]{\@secondoftwo}%
\providecommand \bibfield  [0]{\@secondoftwo}%
\providecommand \translation [1]{[#1]}%
\providecommand \BibitemOpen [0]{}%
\providecommand \bibitemStop [0]{}%
\providecommand \bibitemNoStop [0]{.\EOS\space}%
\providecommand \EOS [0]{\spacefactor3000\relax}%
\providecommand \BibitemShut  [1]{\csname bibitem#1\endcsname}%
\let\auto@bib@innerbib\@empty
\bibitem [{\citenamefont {Abbott}\ \emph {et~al.}(2016)\citenamefont {Abbott}
  \emph {et~al.}}]{TheLIGOScientific:2016pea}%
  \BibitemOpen
  \bibfield  {author} {\bibinfo {author} {\bibfnamefont {B.~P.}\ \bibnamefont
  {Abbott}} \emph {et~al.} (\bibinfo {collaboration} {LIGO Scientific,
  Virgo}),\ }\href {\doibase 10.1103/PhysRevX.6.041015,
  10.1103/PhysRevX.8.039903} {\bibfield  {journal} {\bibinfo  {journal} {\emph
  {Phys. Rev. X}}\ }\textbf {\bibinfo {volume} {6}},\ \bibinfo {pages} {041015}
  (\bibinfo {year} {2016})},\ \bibinfo {note} {[erratum: Phys.
  Rev.X8,no.3,039903(2018)]},\ \Eprint {http://arxiv.org/abs/1606.04856}
  {arXiv:1606.04856 [gr-qc]} \BibitemShut {NoStop}%
\bibitem [{\citenamefont {Akiyama}\ \emph {et~al.}(2019)\citenamefont {Akiyama}
  \emph {et~al.}}]{Akiyama:2019cqa}%
  \BibitemOpen
  \bibfield  {author} {\bibinfo {author} {\bibfnamefont {K.}~\bibnamefont
  {Akiyama}} \emph {et~al.} (\bibinfo {collaboration} {Event Horizon
  Telescope}),\ }\href {\doibase 10.3847/2041-8213/ab0ec7} {\bibfield
  {journal} {\bibinfo  {journal} {\emph {Astrophys. J.}}\ }\textbf {\bibinfo
  {volume} {875}},\ \bibinfo {pages} {L1} (\bibinfo {year} {2019})},\ \Eprint
  {http://arxiv.org/abs/1906.11238} {arXiv:1906.11238 [astro-ph.GA]}
  \BibitemShut {NoStop}%
\bibitem [{\citenamefont {Lovelock}(1971)}]{Lovelock:1971yv}%
  \BibitemOpen
  \bibfield  {author} {\bibinfo {author} {\bibfnamefont {D.}~\bibnamefont
  {Lovelock}},\ }\href {\doibase 10.1063/1.1665613} {\bibfield  {journal}
  {\bibinfo  {journal} {\emph {J. Math. Phys.}}\ }\textbf {\bibinfo {volume}
  {12}},\ \bibinfo {pages} {498} (\bibinfo {year} {1971})}\BibitemShut
  {NoStop}%
\bibitem [{\citenamefont {Woodard}(2015)}]{Woodard:2015zca}%
  \BibitemOpen
  \bibfield  {author} {\bibinfo {author} {\bibfnamefont {R.~P.}\ \bibnamefont
  {Woodard}},\ }\href {\doibase 10.4249/scholarpedia.32243} {\bibfield
  {journal} {\bibinfo  {journal} {\emph {Scholarpedia}}\ }\textbf {\bibinfo
  {volume} {10}},\ \bibinfo {pages} {32243} (\bibinfo {year} {2015})},\ \Eprint
  {http://arxiv.org/abs/1506.02210} {arXiv:1506.02210 [hep-th]} \BibitemShut
  {NoStop}%
\bibitem [{\citenamefont {Motohashi}\ and\ \citenamefont
  {Suyama}(2020)}]{Motohashi:2020psc}%
  \BibitemOpen
  \bibfield  {author} {\bibinfo {author} {\bibfnamefont {H.}~\bibnamefont
  {Motohashi}} and \bibinfo {author} {\bibfnamefont {T.}~\bibnamefont
  {Suyama}},\ }\Eprint {http://arxiv.org/abs/2001.02483} {arXiv:2001.02483
  [hep-th]} \BibitemShut {NoStop}%
\bibitem [{\citenamefont {Motohashi}\ and\ \citenamefont
  {Suyama}(2015)}]{Motohashi:2014opa}%
  \BibitemOpen
  \bibfield  {author} {\bibinfo {author} {\bibfnamefont {H.}~\bibnamefont
  {Motohashi}} and \bibinfo {author} {\bibfnamefont {T.}~\bibnamefont
  {Suyama}},\ }\href {\doibase 10.1103/PhysRevD.91.085009} {\bibfield
  {journal} {\bibinfo  {journal} {\emph {Phys. Rev. D}}\ }\textbf {\bibinfo
  {volume} {91}},\ \bibinfo {pages} {085009} (\bibinfo {year} {2015})},\
  \Eprint {http://arxiv.org/abs/1411.3721} {arXiv:1411.3721 [physics.class-ph]}
  \BibitemShut {NoStop}%
\bibitem [{\citenamefont {Langlois}\ and\ \citenamefont
  {Noui}(2016)}]{Langlois:2015cwa}%
  \BibitemOpen
  \bibfield  {author} {\bibinfo {author} {\bibfnamefont {D.}~\bibnamefont
  {Langlois}} and \bibinfo {author} {\bibfnamefont {K.}~\bibnamefont {Noui}},\
  }\href {\doibase 10.1088/1475-7516/2016/02/034} {\bibfield  {journal}
  {\bibinfo  {journal} {\emph {JCAP}}\ }\textbf {\bibinfo {volume} {02}},\
  \bibinfo {pages} {034} (\bibinfo {year} {2016})},\ \Eprint
  {http://arxiv.org/abs/1510.06930} {arXiv:1510.06930 [gr-qc]} \BibitemShut
  {NoStop}%
\bibitem [{\citenamefont {Motohashi}\ \emph
  {et~al.}(2016{\natexlab{a}})\citenamefont {Motohashi}, \citenamefont {Noui},
  \citenamefont {Suyama}, \citenamefont {Yamaguchi},\ and\ \citenamefont
  {Langlois}}]{Motohashi:2016ftl}%
  \BibitemOpen
  \bibfield  {author} {\bibinfo {author} {\bibfnamefont {H.}~\bibnamefont
  {Motohashi}}, \bibinfo {author} {\bibfnamefont {K.}~\bibnamefont {Noui}},
  \bibinfo {author} {\bibfnamefont {T.}~\bibnamefont {Suyama}}, \bibinfo
  {author} {\bibfnamefont {M.}~\bibnamefont {Yamaguchi}},  and \bibinfo
  {author} {\bibfnamefont {D.}~\bibnamefont {Langlois}},\ }\href {\doibase
  10.1088/1475-7516/2016/07/033} {\bibfield  {journal} {\bibinfo  {journal}
  {\emph {JCAP}}\ }\textbf {\bibinfo {volume} {07}},\ \bibinfo {pages} {033}
  (\bibinfo {year} {2016}{\natexlab{a}})},\ \Eprint
  {http://arxiv.org/abs/1603.09355} {arXiv:1603.09355 [hep-th]} \BibitemShut
  {NoStop}%
\bibitem [{\citenamefont {Klein}\ and\ \citenamefont
  {Roest}(2016)}]{Klein:2016aiq}%
  \BibitemOpen
  \bibfield  {author} {\bibinfo {author} {\bibfnamefont {R.}~\bibnamefont
  {Klein}} and \bibinfo {author} {\bibfnamefont {D.}~\bibnamefont {Roest}},\
  }\href {\doibase 10.1007/JHEP07(2016)130} {\bibfield  {journal} {\bibinfo
  {journal} {\emph {JHEP}}\ }\textbf {\bibinfo {volume} {07}},\ \bibinfo
  {pages} {130} (\bibinfo {year} {2016})},\ \Eprint
  {http://arxiv.org/abs/1604.01719} {arXiv:1604.01719 [hep-th]} \BibitemShut
  {NoStop}%
\bibitem [{\citenamefont {Motohashi}\ \emph
  {et~al.}(2018{\natexlab{a}})\citenamefont {Motohashi}, \citenamefont
  {Suyama},\ and\ \citenamefont {Yamaguchi}}]{Motohashi:2017eya}%
  \BibitemOpen
  \bibfield  {author} {\bibinfo {author} {\bibfnamefont {H.}~\bibnamefont
  {Motohashi}}, \bibinfo {author} {\bibfnamefont {T.}~\bibnamefont {Suyama}},
  and \bibinfo {author} {\bibfnamefont {M.}~\bibnamefont {Yamaguchi}},\ }\href
  {\doibase 10.7566/JPSJ.87.063401} {\bibfield  {journal} {\bibinfo  {journal}
  {\emph {J. Phys. Soc. Jap.}}\ }\textbf {\bibinfo {volume} {87}},\ \bibinfo
  {pages} {063401} (\bibinfo {year} {2018}{\natexlab{a}})},\ \Eprint
  {http://arxiv.org/abs/1711.08125} {arXiv:1711.08125 [hep-th]} \BibitemShut
  {NoStop}%
\bibitem [{\citenamefont {Motohashi}\ \emph
  {et~al.}(2018{\natexlab{b}})\citenamefont {Motohashi}, \citenamefont
  {Suyama},\ and\ \citenamefont {Yamaguchi}}]{Motohashi:2018pxg}%
  \BibitemOpen
  \bibfield  {author} {\bibinfo {author} {\bibfnamefont {H.}~\bibnamefont
  {Motohashi}}, \bibinfo {author} {\bibfnamefont {T.}~\bibnamefont {Suyama}},
  and \bibinfo {author} {\bibfnamefont {M.}~\bibnamefont {Yamaguchi}},\ }\href
  {\doibase 10.1007/JHEP06(2018)133} {\bibfield  {journal} {\bibinfo  {journal}
  {\emph {JHEP}}\ }\textbf {\bibinfo {volume} {06}},\ \bibinfo {pages} {133}
  (\bibinfo {year} {2018}{\natexlab{b}})},\ \Eprint
  {http://arxiv.org/abs/1804.07990} {arXiv:1804.07990 [hep-th]} \BibitemShut
  {NoStop}%
\bibitem [{\citenamefont {Sotiriou}\ and\ \citenamefont
  {Faraoni}(2010)}]{Sotiriou:2008rp}%
  \BibitemOpen
  \bibfield  {author} {\bibinfo {author} {\bibfnamefont {T.~P.}\ \bibnamefont
  {Sotiriou}} and \bibinfo {author} {\bibfnamefont {V.}~\bibnamefont
  {Faraoni}},\ }\href {\doibase 10.1103/RevModPhys.82.451} {\bibfield
  {journal} {\bibinfo  {journal} {\emph {Rev. Mod. Phys.}}\ }\textbf {\bibinfo
  {volume} {82}},\ \bibinfo {pages} {451} (\bibinfo {year} {2010})},\ \Eprint
  {http://arxiv.org/abs/0805.1726} {arXiv:0805.1726 [gr-qc]} \BibitemShut
  {NoStop}%
\bibitem [{\citenamefont {De~Felice}\ and\ \citenamefont
  {Tsujikawa}(2010)}]{DeFelice:2010aj}%
  \BibitemOpen
  \bibfield  {author} {\bibinfo {author} {\bibfnamefont {A.}~\bibnamefont
  {De~Felice}} and \bibinfo {author} {\bibfnamefont {S.}~\bibnamefont
  {Tsujikawa}},\ }\href {\doibase 10.12942/lrr-2010-3} {\bibfield  {journal}
  {\bibinfo  {journal} {\emph {Living Rev. Rel.}}\ }\textbf {\bibinfo {volume}
  {13}},\ \bibinfo {pages} {3} (\bibinfo {year} {2010})},\ \Eprint
  {http://arxiv.org/abs/1002.4928} {arXiv:1002.4928 [gr-qc]} \BibitemShut
  {NoStop}%
\bibitem [{\citenamefont {Horndeski}(1974)}]{Horndeski:1974wa}%
  \BibitemOpen
  \bibfield  {author} {\bibinfo {author} {\bibfnamefont {G.~W.}\ \bibnamefont
  {Horndeski}},\ }\href {\doibase 10.1007/BF01807638} {\bibfield  {journal}
  {\bibinfo  {journal} {\emph {Int. J. Theor. Phys.}}\ }\textbf {\bibinfo
  {volume} {10}},\ \bibinfo {pages} {363} (\bibinfo {year} {1974})}\BibitemShut
  {NoStop}%
\bibitem [{\citenamefont {Deffayet}\ \emph {et~al.}(2011)\citenamefont
  {Deffayet}, \citenamefont {Gao}, \citenamefont {Steer},\ and\ \citenamefont
  {Zahariade}}]{Deffayet:2011gz}%
  \BibitemOpen
  \bibfield  {author} {\bibinfo {author} {\bibfnamefont {C.}~\bibnamefont
  {Deffayet}}, \bibinfo {author} {\bibfnamefont {X.}~\bibnamefont {Gao}},
  \bibinfo {author} {\bibfnamefont {D.~A.}\ \bibnamefont {Steer}},  and
  \bibinfo {author} {\bibfnamefont {G.}~\bibnamefont {Zahariade}},\ }\href
  {\doibase 10.1103/PhysRevD.84.064039} {\bibfield  {journal} {\bibinfo
  {journal} {\emph {Phys. Rev. D}}\ }\textbf {\bibinfo {volume} {84}},\
  \bibinfo {pages} {064039} (\bibinfo {year} {2011})},\ \Eprint
  {http://arxiv.org/abs/1103.3260} {arXiv:1103.3260 [hep-th]} \BibitemShut
  {NoStop}%
\bibitem [{\citenamefont {Kobayashi}\ \emph {et~al.}(2011)\citenamefont
  {Kobayashi}, \citenamefont {Yamaguchi},\ and\ \citenamefont
  {Yokoyama}}]{Kobayashi:2011nu}%
  \BibitemOpen
  \bibfield  {author} {\bibinfo {author} {\bibfnamefont {T.}~\bibnamefont
  {Kobayashi}}, \bibinfo {author} {\bibfnamefont {M.}~\bibnamefont
  {Yamaguchi}},  and \bibinfo {author} {\bibfnamefont {J.}~\bibnamefont
  {Yokoyama}},\ }\href {\doibase 10.1143/PTP.126.511} {\bibfield  {journal}
  {\bibinfo  {journal} {\emph {Prog. Theor. Phys.}}\ }\textbf {\bibinfo
  {volume} {126}},\ \bibinfo {pages} {511} (\bibinfo {year} {2011})},\ \Eprint
  {http://arxiv.org/abs/1105.5723} {arXiv:1105.5723 [hep-th]} \BibitemShut
  {NoStop}%
\bibitem [{\citenamefont {Gleyzes}\ \emph {et~al.}(2015)\citenamefont
  {Gleyzes}, \citenamefont {Langlois}, \citenamefont {Piazza},\ and\
  \citenamefont {Vernizzi}}]{Gleyzes:2014dya}%
  \BibitemOpen
  \bibfield  {author} {\bibinfo {author} {\bibfnamefont {J.}~\bibnamefont
  {Gleyzes}}, \bibinfo {author} {\bibfnamefont {D.}~\bibnamefont {Langlois}},
  \bibinfo {author} {\bibfnamefont {F.}~\bibnamefont {Piazza}},  and \bibinfo
  {author} {\bibfnamefont {F.}~\bibnamefont {Vernizzi}},\ }\href {\doibase
  10.1103/PhysRevLett.114.211101} {\bibfield  {journal} {\bibinfo  {journal}
  {\emph {Phys. Rev. Lett.}}\ }\textbf {\bibinfo {volume} {114}},\ \bibinfo
  {pages} {211101} (\bibinfo {year} {2015})},\ \Eprint
  {http://arxiv.org/abs/1404.6495} {arXiv:1404.6495 [hep-th]} \BibitemShut
  {NoStop}%
\bibitem [{\citenamefont {Crisostomi}\ \emph {et~al.}(2016)\citenamefont
  {Crisostomi}, \citenamefont {Koyama},\ and\ \citenamefont
  {Tasinato}}]{Crisostomi:2016czh}%
  \BibitemOpen
  \bibfield  {author} {\bibinfo {author} {\bibfnamefont {M.}~\bibnamefont
  {Crisostomi}}, \bibinfo {author} {\bibfnamefont {K.}~\bibnamefont {Koyama}},
  and \bibinfo {author} {\bibfnamefont {G.}~\bibnamefont {Tasinato}},\ }\href
  {\doibase 10.1088/1475-7516/2016/04/044} {\bibfield  {journal} {\bibinfo
  {journal} {\emph {JCAP}}\ }\textbf {\bibinfo {volume} {04}},\ \bibinfo
  {pages} {044} (\bibinfo {year} {2016})},\ \Eprint
  {http://arxiv.org/abs/1602.03119} {arXiv:1602.03119 [hep-th]} \BibitemShut
  {NoStop}%
\bibitem [{\citenamefont {Ben~Achour}\ \emph {et~al.}(2016)\citenamefont
  {Ben~Achour}, \citenamefont {Crisostomi}, \citenamefont {Koyama},
  \citenamefont {Langlois}, \citenamefont {Noui},\ and\ \citenamefont
  {Tasinato}}]{BenAchour:2016fzp}%
  \BibitemOpen
  \bibfield  {author} {\bibinfo {author} {\bibfnamefont {J.}~\bibnamefont
  {Ben~Achour}}, \bibinfo {author} {\bibfnamefont {M.}~\bibnamefont
  {Crisostomi}}, \bibinfo {author} {\bibfnamefont {K.}~\bibnamefont {Koyama}},
  \bibinfo {author} {\bibfnamefont {D.}~\bibnamefont {Langlois}}, \bibinfo
  {author} {\bibfnamefont {K.}~\bibnamefont {Noui}},  and \bibinfo {author}
  {\bibfnamefont {G.}~\bibnamefont {Tasinato}},\ }\href {\doibase
  10.1007/JHEP12(2016)100} {\bibfield  {journal} {\bibinfo  {journal} {\emph
  {JHEP}}\ }\textbf {\bibinfo {volume} {12}},\ \bibinfo {pages} {100} (\bibinfo
  {year} {2016})},\ \Eprint {http://arxiv.org/abs/1608.08135} {arXiv:1608.08135
  [hep-th]} \BibitemShut {NoStop}%
\bibitem [{\citenamefont {Takahashi}\ and\ \citenamefont
  {Kobayashi}(2017)}]{Takahashi:2017pje}%
  \BibitemOpen
  \bibfield  {author} {\bibinfo {author} {\bibfnamefont {K.}~\bibnamefont
  {Takahashi}} and \bibinfo {author} {\bibfnamefont {T.}~\bibnamefont
  {Kobayashi}},\ }\href {\doibase 10.1088/1475-7516/2017/11/038} {\bibfield
  {journal} {\bibinfo  {journal} {\emph {JCAP}}\ }\textbf {\bibinfo {volume}
  {11}},\ \bibinfo {pages} {038} (\bibinfo {year} {2017})},\ \Eprint
  {http://arxiv.org/abs/1708.02951} {arXiv:1708.02951 [gr-qc]} \BibitemShut
  {NoStop}%
\bibitem [{\citenamefont {Langlois}\ \emph {et~al.}(2019)\citenamefont
  {Langlois}, \citenamefont {Mancarella}, \citenamefont {Noui},\ and\
  \citenamefont {Vernizzi}}]{Langlois:2018jdg}%
  \BibitemOpen
  \bibfield  {author} {\bibinfo {author} {\bibfnamefont {D.}~\bibnamefont
  {Langlois}}, \bibinfo {author} {\bibfnamefont {M.}~\bibnamefont
  {Mancarella}}, \bibinfo {author} {\bibfnamefont {K.}~\bibnamefont {Noui}},
  and \bibinfo {author} {\bibfnamefont {F.}~\bibnamefont {Vernizzi}},\ }\href
  {\doibase 10.1088/1475-7516/2019/02/036} {\bibfield  {journal} {\bibinfo
  {journal} {\emph {JCAP}}\ }\textbf {\bibinfo {volume} {02}},\ \bibinfo
  {pages} {036} (\bibinfo {year} {2019})},\ \Eprint
  {http://arxiv.org/abs/1802.03394} {arXiv:1802.03394 [gr-qc]} \BibitemShut
  {NoStop}%
\bibitem [{\citenamefont {Langlois}(2019)}]{Langlois:2018dxi}%
  \BibitemOpen
  \bibfield  {author} {\bibinfo {author} {\bibfnamefont {D.}~\bibnamefont
  {Langlois}},\ }\href {\doibase 10.1142/S0218271819420069} {\bibfield
  {journal} {\bibinfo  {journal} {\emph {Int. J. Mod. Phys.}}\ }\textbf
  {\bibinfo {volume} {D28}},\ \bibinfo {pages} {1942006} (\bibinfo {year}
  {2019})},\ \Eprint {http://arxiv.org/abs/1811.06271} {arXiv:1811.06271
  [gr-qc]} \BibitemShut {NoStop}%
\bibitem [{\citenamefont {Kobayashi}(2019)}]{Kobayashi:2019hrl}%
  \BibitemOpen
  \bibfield  {author} {\bibinfo {author} {\bibfnamefont {T.}~\bibnamefont
  {Kobayashi}},\ }\href {\doibase 10.1088/1361-6633/ab2429} {\bibfield
  {journal} {\bibinfo  {journal} {\emph {Rept. Prog. Phys.}}\ }\textbf
  {\bibinfo {volume} {82}},\ \bibinfo {pages} {086901} (\bibinfo {year}
  {2019})},\ \Eprint {http://arxiv.org/abs/1901.07183} {arXiv:1901.07183
  [gr-qc]} \BibitemShut {NoStop}%
\bibitem [{\citenamefont {Lin}\ and\ \citenamefont
  {Mukohyama}(2017)}]{Lin:2017oow}%
  \BibitemOpen
  \bibfield  {author} {\bibinfo {author} {\bibfnamefont {C.}~\bibnamefont
  {Lin}} and \bibinfo {author} {\bibfnamefont {S.}~\bibnamefont {Mukohyama}},\
  }\href {\doibase 10.1088/1475-7516/2017/10/033} {\bibfield  {journal}
  {\bibinfo  {journal} {\emph {JCAP}}\ }\textbf {\bibinfo {volume} {10}},\
  \bibinfo {pages} {033} (\bibinfo {year} {2017})},\ \Eprint
  {http://arxiv.org/abs/1708.03757} {arXiv:1708.03757 [gr-qc]} \BibitemShut
  {NoStop}%
\bibitem [{\citenamefont {Chagoya}\ and\ \citenamefont
  {Tasinato}(2019)}]{Chagoya:2018yna}%
  \BibitemOpen
  \bibfield  {author} {\bibinfo {author} {\bibfnamefont {J.}~\bibnamefont
  {Chagoya}} and \bibinfo {author} {\bibfnamefont {G.}~\bibnamefont
  {Tasinato}},\ }\href {\doibase 10.1088/1361-6382/ab0a4b} {\bibfield
  {journal} {\bibinfo  {journal} {\emph {Class. Quant. Grav.}}\ }\textbf
  {\bibinfo {volume} {36}},\ \bibinfo {pages} {075014} (\bibinfo {year}
  {2019})},\ \Eprint {http://arxiv.org/abs/1805.12010} {arXiv:1805.12010
  [hep-th]} \BibitemShut {NoStop}%
\bibitem [{\citenamefont {Aoki}\ \emph {et~al.}(2018)\citenamefont {Aoki},
  \citenamefont {Lin},\ and\ \citenamefont {Mukohyama}}]{Aoki:2018zcv}%
  \BibitemOpen
  \bibfield  {author} {\bibinfo {author} {\bibfnamefont {K.}~\bibnamefont
  {Aoki}}, \bibinfo {author} {\bibfnamefont {C.}~\bibnamefont {Lin}},  and
  \bibinfo {author} {\bibfnamefont {S.}~\bibnamefont {Mukohyama}},\ }\href
  {\doibase 10.1103/PhysRevD.98.044022} {\bibfield  {journal} {\bibinfo
  {journal} {\emph {Phys. Rev. D}}\ }\textbf {\bibinfo {volume} {98}},\
  \bibinfo {pages} {044022} (\bibinfo {year} {2018})},\ \Eprint
  {http://arxiv.org/abs/1804.03902} {arXiv:1804.03902 [gr-qc]} \BibitemShut
  {NoStop}%
\bibitem [{\citenamefont {Afshordi}\ \emph
  {et~al.}(2007{\natexlab{a}})\citenamefont {Afshordi}, \citenamefont {Chung},\
  and\ \citenamefont {Geshnizjani}}]{Afshordi:2006ad}%
  \BibitemOpen
  \bibfield  {author} {\bibinfo {author} {\bibfnamefont {N.}~\bibnamefont
  {Afshordi}}, \bibinfo {author} {\bibfnamefont {D.~J.~H.}\ \bibnamefont
  {Chung}},  and \bibinfo {author} {\bibfnamefont {G.}~\bibnamefont
  {Geshnizjani}},\ }\href {\doibase 10.1103/PhysRevD.75.083513} {\bibfield
  {journal} {\bibinfo  {journal} {\emph {Phys. Rev. D}}\ }\textbf {\bibinfo
  {volume} {75}},\ \bibinfo {pages} {083513} (\bibinfo {year}
  {2007}{\natexlab{a}})},\ \Eprint {http://arxiv.org/abs/hep-th/0609150}
  {arXiv:hep-th/0609150 [hep-th]} \BibitemShut {NoStop}%
\bibitem [{\citenamefont {Iyonaga}\ \emph {et~al.}(2018)\citenamefont
  {Iyonaga}, \citenamefont {Takahashi},\ and\ \citenamefont
  {Kobayashi}}]{Iyonaga:2018vnu}%
  \BibitemOpen
  \bibfield  {author} {\bibinfo {author} {\bibfnamefont {A.}~\bibnamefont
  {Iyonaga}}, \bibinfo {author} {\bibfnamefont {K.}~\bibnamefont {Takahashi}},
  and \bibinfo {author} {\bibfnamefont {T.}~\bibnamefont {Kobayashi}},\ }\href
  {\doibase 10.1088/1475-7516/2018/12/002} {\bibfield  {journal} {\bibinfo
  {journal} {\emph {JCAP}}\ }\textbf {\bibinfo {volume} {12}},\ \bibinfo
  {pages} {002} (\bibinfo {year} {2018})},\ \Eprint
  {http://arxiv.org/abs/1809.10935} {arXiv:1809.10935 [gr-qc]} \BibitemShut
  {NoStop}%
\bibitem [{\citenamefont {Gao}\ and\ \citenamefont {Yao}(2020)}]{Gao:2019twq}%
  \BibitemOpen
  \bibfield  {author} {\bibinfo {author} {\bibfnamefont {X.}~\bibnamefont
  {Gao}} and \bibinfo {author} {\bibfnamefont {Z.-B.}\ \bibnamefont {Yao}},\
  }\href {\doibase 10.1103/PhysRevD.101.064018} {\bibfield  {journal} {\bibinfo
   {journal} {\emph {Phys. Rev. D}}\ }\textbf {\bibinfo {volume} {101}},\
  \bibinfo {pages} {064018} (\bibinfo {year} {2020})},\ \Eprint
  {http://arxiv.org/abs/1910.13995} {arXiv:1910.13995 [gr-qc]} \BibitemShut
  {NoStop}%
\bibitem [{\citenamefont {Aoki}\ \emph {et~al.}(2019)\citenamefont {Aoki},
  \citenamefont {De~Felice}, \citenamefont {Lin}, \citenamefont {Mukohyama},\
  and\ \citenamefont {Oliosi}}]{Aoki:2018brq}%
  \BibitemOpen
  \bibfield  {author} {\bibinfo {author} {\bibfnamefont {K.}~\bibnamefont
  {Aoki}}, \bibinfo {author} {\bibfnamefont {A.}~\bibnamefont {De~Felice}},
  \bibinfo {author} {\bibfnamefont {C.}~\bibnamefont {Lin}}, \bibinfo {author}
  {\bibfnamefont {S.}~\bibnamefont {Mukohyama}},  and \bibinfo {author}
  {\bibfnamefont {M.}~\bibnamefont {Oliosi}},\ }\href {\doibase
  10.1088/1475-7516/2019/01/017} {\bibfield  {journal} {\bibinfo  {journal}
  {\emph {JCAP}}\ }\textbf {\bibinfo {volume} {01}},\ \bibinfo {pages} {017}
  (\bibinfo {year} {2019})},\ \Eprint {http://arxiv.org/abs/1810.01047}
  {arXiv:1810.01047 [gr-qc]} \BibitemShut {NoStop}%
\bibitem [{\citenamefont {Dom\`enech}\ \emph {et~al.}(2015)\citenamefont
  {Dom\`enech}, \citenamefont {Mukohyama}, \citenamefont {Namba}, \citenamefont
  {Naruko}, \citenamefont {Saitou},\ and\ \citenamefont
  {Watanabe}}]{Domenech:2015tca}%
  \BibitemOpen
  \bibfield  {author} {\bibinfo {author} {\bibfnamefont {G.}~\bibnamefont
  {Dom\`enech}}, \bibinfo {author} {\bibfnamefont {S.}~\bibnamefont
  {Mukohyama}}, \bibinfo {author} {\bibfnamefont {R.}~\bibnamefont {Namba}},
  \bibinfo {author} {\bibfnamefont {A.}~\bibnamefont {Naruko}}, \bibinfo
  {author} {\bibfnamefont {R.}~\bibnamefont {Saitou}},  and \bibinfo {author}
  {\bibfnamefont {Y.}~\bibnamefont {Watanabe}},\ }\href {\doibase
  10.1103/PhysRevD.92.084027} {\bibfield  {journal} {\bibinfo  {journal} {\emph
  {Phys. Rev. D}}\ }\textbf {\bibinfo {volume} {92}},\ \bibinfo {pages}
  {084027} (\bibinfo {year} {2015})},\ \Eprint
  {http://arxiv.org/abs/1507.05390} {arXiv:1507.05390 [hep-th]} \BibitemShut
  {NoStop}%
\bibitem [{\citenamefont {Takahashi}\ \emph {et~al.}(2017)\citenamefont
  {Takahashi}, \citenamefont {Motohashi}, \citenamefont {Suyama},\ and\
  \citenamefont {Kobayashi}}]{Takahashi:2017zgr}%
  \BibitemOpen
  \bibfield  {author} {\bibinfo {author} {\bibfnamefont {K.}~\bibnamefont
  {Takahashi}}, \bibinfo {author} {\bibfnamefont {H.}~\bibnamefont
  {Motohashi}}, \bibinfo {author} {\bibfnamefont {T.}~\bibnamefont {Suyama}},
  and \bibinfo {author} {\bibfnamefont {T.}~\bibnamefont {Kobayashi}},\ }\href
  {\doibase 10.1103/PhysRevD.95.084053} {\bibfield  {journal} {\bibinfo
  {journal} {\emph {Phys. Rev. D}}\ }\textbf {\bibinfo {volume} {95}},\
  \bibinfo {pages} {084053} (\bibinfo {year} {2017})},\ \Eprint
  {http://arxiv.org/abs/1702.01849} {arXiv:1702.01849 [gr-qc]} \BibitemShut
  {NoStop}%
\bibitem [{\citenamefont {Gomes}\ and\ \citenamefont
  {Guariento}(2017)}]{Gomes:2017tzd}%
  \BibitemOpen
  \bibfield  {author} {\bibinfo {author} {\bibfnamefont {H.}~\bibnamefont
  {Gomes}} and \bibinfo {author} {\bibfnamefont {D.~C.}\ \bibnamefont
  {Guariento}},\ }\href {\doibase 10.1103/PhysRevD.95.104049} {\bibfield
  {journal} {\bibinfo  {journal} {\emph {Phys. Rev. D}}\ }\textbf {\bibinfo
  {volume} {95}},\ \bibinfo {pages} {104049} (\bibinfo {year} {2017})},\
  \Eprint {http://arxiv.org/abs/1703.08226} {arXiv:1703.08226 [gr-qc]}
  \BibitemShut {NoStop}%
\bibitem [{\citenamefont {Afshordi}\ \emph
  {et~al.}(2007{\natexlab{b}})\citenamefont {Afshordi}, \citenamefont {Chung},
  \citenamefont {Doran},\ and\ \citenamefont {Geshnizjani}}]{Afshordi:2007yx}%
  \BibitemOpen
  \bibfield  {author} {\bibinfo {author} {\bibfnamefont {N.}~\bibnamefont
  {Afshordi}}, \bibinfo {author} {\bibfnamefont {D.~J.~H.}\ \bibnamefont
  {Chung}}, \bibinfo {author} {\bibfnamefont {M.}~\bibnamefont {Doran}},  and
  \bibinfo {author} {\bibfnamefont {G.}~\bibnamefont {Geshnizjani}},\ }\href
  {\doibase 10.1103/PhysRevD.75.123509} {\bibfield  {journal} {\bibinfo
  {journal} {\emph {Phys. Rev. D}}\ }\textbf {\bibinfo {volume} {75}},\
  \bibinfo {pages} {123509} (\bibinfo {year} {2007}{\natexlab{b}})},\ \Eprint
  {http://arxiv.org/abs/astro-ph/0702002} {arXiv:astro-ph/0702002 [astro-ph]}
  \BibitemShut {NoStop}%
\bibitem [{\citenamefont {Boruah}\ \emph {et~al.}(2018)\citenamefont {Boruah},
  \citenamefont {Kim}, \citenamefont {Rouben},\ and\ \citenamefont
  {Geshnizjani}}]{Boruah:2018pvq}%
  \BibitemOpen
  \bibfield  {author} {\bibinfo {author} {\bibfnamefont {S.~S.}\ \bibnamefont
  {Boruah}}, \bibinfo {author} {\bibfnamefont {H.~J.}\ \bibnamefont {Kim}},
  \bibinfo {author} {\bibfnamefont {M.}~\bibnamefont {Rouben}},  and \bibinfo
  {author} {\bibfnamefont {G.}~\bibnamefont {Geshnizjani}},\ }\href {\doibase
  10.1088/1475-7516/2018/08/031} {\bibfield  {journal} {\bibinfo  {journal}
  {\emph {JCAP}}\ }\textbf {\bibinfo {volume} {08}},\ \bibinfo {pages} {031}
  (\bibinfo {year} {2018})},\ \Eprint {http://arxiv.org/abs/1802.06818}
  {arXiv:1802.06818 [gr-qc]} \BibitemShut {NoStop}%
\bibitem [{\citenamefont {Quintin}\ and\ \citenamefont
  {Yoshida}(2020)}]{Quintin:2019orx}%
  \BibitemOpen
  \bibfield  {author} {\bibinfo {author} {\bibfnamefont {J.}~\bibnamefont
  {Quintin}} and \bibinfo {author} {\bibfnamefont {D.}~\bibnamefont
  {Yoshida}},\ }\href {\doibase 10.1088/1475-7516/2020/02/016} {\bibfield
  {journal} {\bibinfo  {journal} {\emph {JCAP}}\ }\textbf {\bibinfo {volume}
  {02}},\ \bibinfo {pages} {016} (\bibinfo {year} {2020})},\ \Eprint
  {http://arxiv.org/abs/1911.06040} {arXiv:1911.06040 [gr-qc]} \BibitemShut
  {NoStop}%
\bibitem [{\citenamefont {Ito}\ \emph {et~al.}(2019{\natexlab{a}})\citenamefont
  {Ito}, \citenamefont {Iyonaga}, \citenamefont {Kim},\ and\ \citenamefont
  {Soda}}]{Ito:2019fie}%
  \BibitemOpen
  \bibfield  {author} {\bibinfo {author} {\bibfnamefont {A.}~\bibnamefont
  {Ito}}, \bibinfo {author} {\bibfnamefont {A.}~\bibnamefont {Iyonaga}},
  \bibinfo {author} {\bibfnamefont {S.}~\bibnamefont {Kim}},  and \bibinfo
  {author} {\bibfnamefont {J.}~\bibnamefont {Soda}},\ }\href {\doibase
  10.1103/PhysRevD.99.083502} {\bibfield  {journal} {\bibinfo  {journal} {\emph
  {Phys. Rev. D}}\ }\textbf {\bibinfo {volume} {99}},\ \bibinfo {pages}
  {083502} (\bibinfo {year} {2019}{\natexlab{a}})},\ \Eprint
  {http://arxiv.org/abs/1902.08663} {arXiv:1902.08663 [astro-ph.CO]}
  \BibitemShut {NoStop}%
\bibitem [{\citenamefont {Ito}\ \emph {et~al.}(2019{\natexlab{b}})\citenamefont
  {Ito}, \citenamefont {Sakakihara},\ and\ \citenamefont {Soda}}]{Ito:2019ztb}%
  \BibitemOpen
  \bibfield  {author} {\bibinfo {author} {\bibfnamefont {A.}~\bibnamefont
  {Ito}}, \bibinfo {author} {\bibfnamefont {Y.}~\bibnamefont {Sakakihara}},
  and \bibinfo {author} {\bibfnamefont {J.}~\bibnamefont {Soda}},\ }\href
  {\doibase 10.1103/PhysRevD.100.063531} {\bibfield  {journal} {\bibinfo
  {journal} {\emph {Phys. Rev. D}}\ }\textbf {\bibinfo {volume} {100}},\
  \bibinfo {pages} {063531} (\bibinfo {year} {2019}{\natexlab{b}})},\ \Eprint
  {http://arxiv.org/abs/1906.10363} {arXiv:1906.10363 [gr-qc]} \BibitemShut
  {NoStop}%
\bibitem [{\citenamefont {Afshordi}(2009)}]{Afshordi:2009tt}%
  \BibitemOpen
  \bibfield  {author} {\bibinfo {author} {\bibfnamefont {N.}~\bibnamefont
  {Afshordi}},\ }\href {\doibase 10.1103/PhysRevD.80.081502} {\bibfield
  {journal} {\bibinfo  {journal} {\emph {Phys. Rev. D}}\ }\textbf {\bibinfo
  {volume} {80}},\ \bibinfo {pages} {081502} (\bibinfo {year} {2009})},\
  \Eprint {http://arxiv.org/abs/0907.5201} {arXiv:0907.5201 [hep-th]}
  \BibitemShut {NoStop}%
\bibitem [{\citenamefont {Bhattacharyya}\ \emph {et~al.}(2018)\citenamefont
  {Bhattacharyya}, \citenamefont {Coates}, \citenamefont {Colombo},
  \citenamefont {G{\"u}mr{\"u}k\c{c}{\"u}o\u{g}lu},\ and\ \citenamefont
  {Sotiriou}}]{Bhattacharyya:2016mah}%
  \BibitemOpen
  \bibfield  {author} {\bibinfo {author} {\bibfnamefont {J.}~\bibnamefont
  {Bhattacharyya}}, \bibinfo {author} {\bibfnamefont {A.}~\bibnamefont
  {Coates}}, \bibinfo {author} {\bibfnamefont {M.}~\bibnamefont {Colombo}},
  \bibinfo {author} {\bibfnamefont {A.~E.}\ \bibnamefont
  {G{\"u}mr{\"u}k\c{c}{\"u}o\u{g}lu}},  and \bibinfo {author} {\bibfnamefont
  {T.~P.}\ \bibnamefont {Sotiriou}},\ }\href {\doibase
  10.1103/PhysRevD.97.064020} {\bibfield  {journal} {\bibinfo  {journal} {\emph
  {Phys. Rev. D}}\ }\textbf {\bibinfo {volume} {97}},\ \bibinfo {pages}
  {064020} (\bibinfo {year} {2018})},\ \Eprint
  {http://arxiv.org/abs/1612.01824} {arXiv:1612.01824 [hep-th]} \BibitemShut
  {NoStop}%
\bibitem [{\citenamefont {de~Rham}\ and\ \citenamefont
  {Motohashi}(2017)}]{deRham:2016ged}%
  \BibitemOpen
  \bibfield  {author} {\bibinfo {author} {\bibfnamefont {C.}~\bibnamefont
  {de~Rham}} and \bibinfo {author} {\bibfnamefont {H.}~\bibnamefont
  {Motohashi}},\ }\href {\doibase 10.1103/PhysRevD.95.064008} {\bibfield
  {journal} {\bibinfo  {journal} {\emph {Phys. Rev. D}}\ }\textbf {\bibinfo
  {volume} {95}},\ \bibinfo {pages} {064008} (\bibinfo {year} {2017})},\
  \Eprint {http://arxiv.org/abs/1611.05038} {arXiv:1611.05038 [hep-th]}
  \BibitemShut {NoStop}%
\bibitem [{\citenamefont {Pajer}\ and\ \citenamefont
  {Stefanyszyn}(2019)}]{Pajer:2018egx}%
  \BibitemOpen
  \bibfield  {author} {\bibinfo {author} {\bibfnamefont {E.}~\bibnamefont
  {Pajer}} and \bibinfo {author} {\bibfnamefont {D.}~\bibnamefont
  {Stefanyszyn}},\ }\href {\doibase 10.1007/JHEP06(2019)008} {\bibfield
  {journal} {\bibinfo  {journal} {\emph {JHEP}}\ }\textbf {\bibinfo {volume}
  {06}},\ \bibinfo {pages} {008} (\bibinfo {year} {2019})},\ \Eprint
  {http://arxiv.org/abs/1812.05133} {arXiv:1812.05133 [hep-th]} \BibitemShut
  {NoStop}%
\bibitem [{\citenamefont {Grall}\ \emph {et~al.}(2020)\citenamefont {Grall},
  \citenamefont {Jazayeri},\ and\ \citenamefont {Pajer}}]{Grall:2019qof}%
  \BibitemOpen
  \bibfield  {author} {\bibinfo {author} {\bibfnamefont {T.}~\bibnamefont
  {Grall}}, \bibinfo {author} {\bibfnamefont {S.}~\bibnamefont {Jazayeri}},
  and \bibinfo {author} {\bibfnamefont {E.}~\bibnamefont {Pajer}},\ }\href
  {\doibase 10.1088/1475-7516/2020/05/031} {\bibfield  {journal} {\bibinfo
  {journal} {\emph {JCAP}}\ }\textbf {\bibinfo {volume} {05}},\ \bibinfo
  {pages} {031} (\bibinfo {year} {2020})},\ \Eprint
  {http://arxiv.org/abs/1909.04622} {arXiv:1909.04622 [hep-th]} \BibitemShut
  {NoStop}%
\bibitem [{\citenamefont {Abbott}\ \emph
  {et~al.}(2017{\natexlab{a}})\citenamefont {Abbott} \emph
  {et~al.}}]{TheLIGOScientific:2017qsa}%
  \BibitemOpen
  \bibfield  {author} {\bibinfo {author} {\bibfnamefont {B.~P.}\ \bibnamefont
  {Abbott}} \emph {et~al.},\ }\href {\doibase 10.1103/PhysRevLett.119.161101}
  {\bibfield  {journal} {\bibinfo  {journal} {\emph {Phys. Rev. Lett.}}\
  }\textbf {\bibinfo {volume} {119}},\ \bibinfo {pages} {161101} (\bibinfo
  {year} {2017}{\natexlab{a}})},\ \Eprint {http://arxiv.org/abs/1710.05832}
  {arXiv:1710.05832 [gr-qc]} \BibitemShut {NoStop}%
\bibitem [{\citenamefont {Abbott}\ \emph
  {et~al.}(2017{\natexlab{b}})\citenamefont {Abbott} \emph
  {et~al.}}]{GBM:2017lvd}%
  \BibitemOpen
  \bibfield  {author} {\bibinfo {author} {\bibfnamefont {B.~P.}\ \bibnamefont
  {Abbott}} \emph {et~al.},\ }\href {\doibase 10.3847/2041-8213/aa91c9}
  {\bibfield  {journal} {\bibinfo  {journal} {\emph {Astrophys. J.}}\ }\textbf
  {\bibinfo {volume} {848}},\ \bibinfo {pages} {L12} (\bibinfo {year}
  {2017}{\natexlab{b}})},\ \Eprint {http://arxiv.org/abs/1710.05833}
  {arXiv:1710.05833 [astro-ph.HE]} \BibitemShut {NoStop}%
\bibitem [{\citenamefont {Abbott}\ \emph
  {et~al.}(2017{\natexlab{c}})\citenamefont {Abbott} \emph
  {et~al.}}]{Monitor:2017mdv}%
  \BibitemOpen
  \bibfield  {author} {\bibinfo {author} {\bibfnamefont {B.~P.}\ \bibnamefont
  {Abbott}} \emph {et~al.},\ }\href {\doibase 10.3847/2041-8213/aa920c}
  {\bibfield  {journal} {\bibinfo  {journal} {\emph {Astrophys. J.}}\ }\textbf
  {\bibinfo {volume} {848}},\ \bibinfo {pages} {L13} (\bibinfo {year}
  {2017}{\natexlab{c}})},\ \Eprint {http://arxiv.org/abs/1710.05834}
  {arXiv:1710.05834 [astro-ph.HE]} \BibitemShut {NoStop}%
\bibitem [{\citenamefont {Sakstein}\ and\ \citenamefont
  {Jain}(2017)}]{Sakstein:2017xjx}%
  \BibitemOpen
  \bibfield  {author} {\bibinfo {author} {\bibfnamefont {J.}~\bibnamefont
  {Sakstein}} and \bibinfo {author} {\bibfnamefont {B.}~\bibnamefont {Jain}},\
  }\href {\doibase 10.1103/PhysRevLett.119.251303} {\bibfield  {journal}
  {\bibinfo  {journal} {\emph {Phys. Rev. Lett.}}\ }\textbf {\bibinfo {volume}
  {119}},\ \bibinfo {pages} {251303} (\bibinfo {year} {2017})},\ \Eprint
  {http://arxiv.org/abs/1710.05893} {arXiv:1710.05893 [astro-ph.CO]}
  \BibitemShut {NoStop}%
\bibitem [{\citenamefont {Deffayet}\ \emph {et~al.}(2010)\citenamefont
  {Deffayet}, \citenamefont {Pujol{\`a}s}, \citenamefont {Sawicki},\ and\
  \citenamefont {Vikman}}]{Deffayet:2010qz}%
  \BibitemOpen
  \bibfield  {author} {\bibinfo {author} {\bibfnamefont {C.}~\bibnamefont
  {Deffayet}}, \bibinfo {author} {\bibfnamefont {O.}~\bibnamefont
  {Pujol{\`a}s}}, \bibinfo {author} {\bibfnamefont {I.}~\bibnamefont
  {Sawicki}},  and \bibinfo {author} {\bibfnamefont {A.}~\bibnamefont
  {Vikman}},\ }\href {\doibase 10.1088/1475-7516/2010/10/026} {\bibfield
  {journal} {\bibinfo  {journal} {\emph {JCAP}}\ }\textbf {\bibinfo {volume}
  {10}},\ \bibinfo {pages} {026} (\bibinfo {year} {2010})},\ \Eprint
  {http://arxiv.org/abs/1008.0048} {arXiv:1008.0048 [hep-th]} \BibitemShut
  {NoStop}%
\bibitem [{\citenamefont {Kobayashi}\ \emph {et~al.}(2010)\citenamefont
  {Kobayashi}, \citenamefont {Yamaguchi},\ and\ \citenamefont
  {Yokoyama}}]{Kobayashi:2010cm}%
  \BibitemOpen
  \bibfield  {author} {\bibinfo {author} {\bibfnamefont {T.}~\bibnamefont
  {Kobayashi}}, \bibinfo {author} {\bibfnamefont {M.}~\bibnamefont
  {Yamaguchi}},  and \bibinfo {author} {\bibfnamefont {J.}~\bibnamefont
  {Yokoyama}},\ }\href {\doibase 10.1103/PhysRevLett.105.231302} {\bibfield
  {journal} {\bibinfo  {journal} {\emph {Phys. Rev. Lett.}}\ }\textbf {\bibinfo
  {volume} {105}},\ \bibinfo {pages} {231302} (\bibinfo {year} {2010})},\
  \Eprint {http://arxiv.org/abs/1008.0603} {arXiv:1008.0603 [hep-th]}
  \BibitemShut {NoStop}%
\bibitem [{\citenamefont {Pujol{\`a}s}\ \emph {et~al.}(2011)\citenamefont
  {Pujol{\`a}s}, \citenamefont {Sawicki},\ and\ \citenamefont
  {Vikman}}]{Pujolas:2011he}%
  \BibitemOpen
  \bibfield  {author} {\bibinfo {author} {\bibfnamefont {O.}~\bibnamefont
  {Pujol{\`a}s}}, \bibinfo {author} {\bibfnamefont {I.}~\bibnamefont
  {Sawicki}},  and \bibinfo {author} {\bibfnamefont {A.}~\bibnamefont
  {Vikman}},\ }\href {\doibase 10.1007/JHEP11(2011)156} {\bibfield  {journal}
  {\bibinfo  {journal} {\emph {JHEP}}\ }\textbf {\bibinfo {volume} {11}},\
  \bibinfo {pages} {156} (\bibinfo {year} {2011})},\ \Eprint
  {http://arxiv.org/abs/1103.5360} {arXiv:1103.5360 [hep-th]} \BibitemShut
  {NoStop}%
\bibitem [{\citenamefont {Afshordi}\ \emph {et~al.}(2014)\citenamefont
  {Afshordi}, \citenamefont {Fontanini},\ and\ \citenamefont
  {Guariento}}]{Afshordi:2014qaa}%
  \BibitemOpen
  \bibfield  {author} {\bibinfo {author} {\bibfnamefont {N.}~\bibnamefont
  {Afshordi}}, \bibinfo {author} {\bibfnamefont {M.}~\bibnamefont {Fontanini}},
   and \bibinfo {author} {\bibfnamefont {D.~C.}\ \bibnamefont {Guariento}},\
  }\href {\doibase 10.1103/PhysRevD.90.084012} {\bibfield  {journal} {\bibinfo
  {journal} {\emph {Phys. Rev. D}}\ }\textbf {\bibinfo {volume} {90}},\
  \bibinfo {pages} {084012} (\bibinfo {year} {2014})},\ \Eprint
  {http://arxiv.org/abs/1408.5538} {arXiv:1408.5538 [gr-qc]} \BibitemShut
  {NoStop}%
\bibitem [{\citenamefont {Gao}(2014)}]{Gao:2014soa}%
  \BibitemOpen
  \bibfield  {author} {\bibinfo {author} {\bibfnamefont {X.}~\bibnamefont
  {Gao}},\ }\href {\doibase 10.1103/PhysRevD.90.081501} {\bibfield  {journal}
  {\bibinfo  {journal} {\emph {Phys. Rev. D}}\ }\textbf {\bibinfo {volume}
  {90}},\ \bibinfo {pages} {081501} (\bibinfo {year} {2014})},\ \Eprint
  {http://arxiv.org/abs/1406.0822} {arXiv:1406.0822 [gr-qc]} \BibitemShut
  {NoStop}%
\bibitem [{\citenamefont {De~Felice}\ \emph {et~al.}(2018)\citenamefont
  {De~Felice}, \citenamefont {Langlois}, \citenamefont {Mukohyama},
  \citenamefont {Noui},\ and\ \citenamefont {Wang}}]{DeFelice:2018mkq}%
  \BibitemOpen
  \bibfield  {author} {\bibinfo {author} {\bibfnamefont {A.}~\bibnamefont
  {De~Felice}}, \bibinfo {author} {\bibfnamefont {D.}~\bibnamefont {Langlois}},
  \bibinfo {author} {\bibfnamefont {S.}~\bibnamefont {Mukohyama}}, \bibinfo
  {author} {\bibfnamefont {K.}~\bibnamefont {Noui}},  and \bibinfo {author}
  {\bibfnamefont {A.}~\bibnamefont {Wang}},\ }\href {\doibase
  10.1103/PhysRevD.98.084024} {\bibfield  {journal} {\bibinfo  {journal} {\emph
  {Phys. Rev.D}}\ }\textbf {\bibinfo {volume} {98}},\ \bibinfo {pages} {084024}
  (\bibinfo {year} {2018})},\ \Eprint {http://arxiv.org/abs/1803.06241}
  {arXiv:1803.06241 [hep-th]} \BibitemShut {NoStop}%
\bibitem [{\citenamefont {Blas}\ \emph {et~al.}(2011)\citenamefont {Blas},
  \citenamefont {Pujol\`as},\ and\ \citenamefont {Sibiryakov}}]{Blas:2010hb}%
  \BibitemOpen
  \bibfield  {author} {\bibinfo {author} {\bibfnamefont {D.}~\bibnamefont
  {Blas}}, \bibinfo {author} {\bibfnamefont {O.}~\bibnamefont {Pujol\`as}},
  and \bibinfo {author} {\bibfnamefont {S.}~\bibnamefont {Sibiryakov}},\ }\href
  {\doibase 10.1007/JHEP04(2011)018} {\bibfield  {journal} {\bibinfo  {journal}
  {\emph {JHEP}}\ }\textbf {\bibinfo {volume} {04}},\ \bibinfo {pages} {018}
  (\bibinfo {year} {2011})},\ \Eprint {http://arxiv.org/abs/1007.3503}
  {arXiv:1007.3503 [hep-th]} \BibitemShut {NoStop}%
\bibitem [{\citenamefont {Will}(2014)}]{Will:2014kxa}%
  \BibitemOpen
  \bibfield  {author} {\bibinfo {author} {\bibfnamefont {C.~M.}\ \bibnamefont
  {Will}},\ }\href {\doibase 10.12942/lrr-2014-4} {\bibfield  {journal}
  {\bibinfo  {journal} {\emph {Living Rev. Rel.}}\ }\textbf {\bibinfo {volume}
  {17}},\ \bibinfo {pages} {4} (\bibinfo {year} {2014})},\ \Eprint
  {http://arxiv.org/abs/1403.7377} {arXiv:1403.7377 [gr-qc]} \BibitemShut
  {NoStop}%
\bibitem [{\citenamefont {Emir~G{\"u}mr{\"u}k\c{c}{\"u}o\u{g}lu}\ \emph
  {et~al.}(2018)\citenamefont {Emir~G{\"u}mr{\"u}k\c{c}{\"u}o\u{g}lu},
  \citenamefont {Saravani},\ and\ \citenamefont
  {Sotiriou}}]{Gumrukcuoglu:2017ijh}%
  \BibitemOpen
  \bibfield  {author} {\bibinfo {author} {\bibfnamefont {A.}~\bibnamefont
  {Emir~G{\"u}mr{\"u}k\c{c}{\"u}o\u{g}lu}}, \bibinfo {author} {\bibfnamefont
  {M.}~\bibnamefont {Saravani}},  and \bibinfo {author} {\bibfnamefont {T.~P.}\
  \bibnamefont {Sotiriou}},\ }\href {\doibase 10.1103/PhysRevD.97.024032}
  {\bibfield  {journal} {\bibinfo  {journal} {\emph {Phys. Rev. D}}\ }\textbf
  {\bibinfo {volume} {97}},\ \bibinfo {pages} {024032} (\bibinfo {year}
  {2018})},\ \Eprint {http://arxiv.org/abs/1711.08845} {arXiv:1711.08845
  [gr-qc]} \BibitemShut {NoStop}%
\bibitem [{\citenamefont {Ramos}\ and\ \citenamefont
  {Barausse}(2019)}]{Ramos:2018oku}%
  \BibitemOpen
  \bibfield  {author} {\bibinfo {author} {\bibfnamefont {O.}~\bibnamefont
  {Ramos}} and \bibinfo {author} {\bibfnamefont {E.}~\bibnamefont {Barausse}},\
  }\href {\doibase 10.1103/PhysRevD.99.024034} {\bibfield  {journal} {\bibinfo
  {journal} {\emph {Phys. Rev. D}}\ }\textbf {\bibinfo {volume} {99}},\
  \bibinfo {pages} {024034} (\bibinfo {year} {2019})},\ \Eprint
  {http://arxiv.org/abs/1811.07786} {arXiv:1811.07786 [gr-qc]} \BibitemShut
  {NoStop}%
\bibitem [{\citenamefont {Blas}\ \emph {et~al.}(2009)\citenamefont {Blas},
  \citenamefont {Pujol\`as},\ and\ \citenamefont {Sibiryakov}}]{Blas:2009yd}%
  \BibitemOpen
  \bibfield  {author} {\bibinfo {author} {\bibfnamefont {D.}~\bibnamefont
  {Blas}}, \bibinfo {author} {\bibfnamefont {O.}~\bibnamefont {Pujol\`as}},
  and \bibinfo {author} {\bibfnamefont {S.}~\bibnamefont {Sibiryakov}},\ }\href
  {\doibase 10.1088/1126-6708/2009/10/029} {\bibfield  {journal} {\bibinfo
  {journal} {\emph {JHEP}}\ }\textbf {\bibinfo {volume} {10}},\ \bibinfo
  {pages} {029} (\bibinfo {year} {2009})},\ \Eprint
  {http://arxiv.org/abs/0906.3046} {arXiv:0906.3046 [hep-th]} \BibitemShut
  {NoStop}%
\bibitem [{\citenamefont {Armend\'ariz-Pic\'on}\ \emph
  {et~al.}(1999)\citenamefont {Armend\'ariz-Pic\'on}, \citenamefont {Damour},\
  and\ \citenamefont {Mukhanov}}]{ArmendarizPicon:1999rj}%
  \BibitemOpen
  \bibfield  {author} {\bibinfo {author} {\bibfnamefont {C.}~\bibnamefont
  {Armend\'ariz-Pic\'on}}, \bibinfo {author} {\bibfnamefont {T.}~\bibnamefont
  {Damour}},  and \bibinfo {author} {\bibfnamefont {V.~F.}\ \bibnamefont
  {Mukhanov}},\ }\href {\doibase 10.1016/S0370-2693(99)00603-6} {\bibfield
  {journal} {\bibinfo  {journal} {\emph {Phys. Lett. B}}\ }\textbf {\bibinfo
  {volume} {458}},\ \bibinfo {pages} {209} (\bibinfo {year} {1999})},\ \Eprint
  {http://arxiv.org/abs/hep-th/9904075} {arXiv:hep-th/9904075 [hep-th]}
  \BibitemShut {NoStop}%
\bibitem [{\citenamefont {Motohashi}\ \emph
  {et~al.}(2016{\natexlab{b}})\citenamefont {Motohashi}, \citenamefont
  {Suyama},\ and\ \citenamefont {Takahashi}}]{Motohashi:2016prk}%
  \BibitemOpen
  \bibfield  {author} {\bibinfo {author} {\bibfnamefont {H.}~\bibnamefont
  {Motohashi}}, \bibinfo {author} {\bibfnamefont {T.}~\bibnamefont {Suyama}},
  and \bibinfo {author} {\bibfnamefont {K.}~\bibnamefont {Takahashi}},\ }\href
  {\doibase 10.1103/PhysRevD.94.124021} {\bibfield  {journal} {\bibinfo
  {journal} {\emph {Phys. Rev. D}}\ }\textbf {\bibinfo {volume} {94}},\
  \bibinfo {pages} {124021} (\bibinfo {year} {2016}{\natexlab{b}})},\ \Eprint
  {http://arxiv.org/abs/1608.00071} {arXiv:1608.00071 [gr-qc]} \BibitemShut
  {NoStop}%
\bibitem [{\citenamefont {Boubekeur}\ \emph {et~al.}(2008)\citenamefont
  {Boubekeur}, \citenamefont {Creminelli}, \citenamefont {Nore{\~n}a},\ and\
  \citenamefont {Vernizzi}}]{Boubekeur:2008kn}%
  \BibitemOpen
  \bibfield  {author} {\bibinfo {author} {\bibfnamefont {L.}~\bibnamefont
  {Boubekeur}}, \bibinfo {author} {\bibfnamefont {P.}~\bibnamefont
  {Creminelli}}, \bibinfo {author} {\bibfnamefont {J.}~\bibnamefont
  {Nore{\~n}a}},  and \bibinfo {author} {\bibfnamefont {F.}~\bibnamefont
  {Vernizzi}},\ }\href {\doibase 10.1088/1475-7516/2008/08/028} {\bibfield
  {journal} {\bibinfo  {journal} {\emph {JCAP}}\ }\textbf {\bibinfo {volume}
  {08}},\ \bibinfo {pages} {028} (\bibinfo {year} {2008})},\ \Eprint
  {http://arxiv.org/abs/0806.1016} {arXiv:0806.1016 [astro-ph]} \BibitemShut
  {NoStop}%
\bibitem [{\citenamefont {De~Felice}\ and\ \citenamefont
  {Mukohyama}(2016)}]{DeFelice:2015moy}%
  \BibitemOpen
  \bibfield  {author} {\bibinfo {author} {\bibfnamefont {A.}~\bibnamefont
  {De~Felice}} and \bibinfo {author} {\bibfnamefont {S.}~\bibnamefont
  {Mukohyama}},\ }\href {\doibase 10.1088/1475-7516/2016/04/028} {\bibfield
  {journal} {\bibinfo  {journal} {\emph {JCAP}}\ }\textbf {\bibinfo {volume}
  {04}},\ \bibinfo {pages} {028} (\bibinfo {year} {2016})},\ \Eprint
  {http://arxiv.org/abs/1512.04008} {arXiv:1512.04008 [hep-th]} \BibitemShut
  {NoStop}%
\bibitem [{\citenamefont {Babichev}\ \emph {et~al.}(2018)\citenamefont
  {Babichev}, \citenamefont {Ramazanov},\ and\ \citenamefont
  {Vikman}}]{Babichev:2018twg}%
  \BibitemOpen
  \bibfield  {author} {\bibinfo {author} {\bibfnamefont {E.}~\bibnamefont
  {Babichev}}, \bibinfo {author} {\bibfnamefont {S.}~\bibnamefont {Ramazanov}},
   and \bibinfo {author} {\bibfnamefont {A.}~\bibnamefont {Vikman}},\ }\href
  {\doibase 10.1088/1475-7516/2018/11/023} {\bibfield  {journal} {\bibinfo
  {journal} {\emph {JCAP}}\ }\textbf {\bibinfo {volume} {11}},\ \bibinfo
  {pages} {023} (\bibinfo {year} {2018})},\ \Eprint
  {http://arxiv.org/abs/1807.10281} {arXiv:1807.10281 [gr-qc]} \BibitemShut
  {NoStop}%
\bibitem [{\citenamefont {Brown}\ and\ \citenamefont {Kucha{\v
  r}}(1995)}]{Brown:1994py}%
  \BibitemOpen
  \bibfield  {author} {\bibinfo {author} {\bibfnamefont {J.}~\bibnamefont
  {Brown}} and \bibinfo {author} {\bibfnamefont {K.~V.}\ \bibnamefont {Kucha{\v
  r}}},\ }\href {\doibase 10.1103/PhysRevD.51.5600} {\bibfield  {journal}
  {\bibinfo  {journal} {\emph {Phys. Rev. D}}\ }\textbf {\bibinfo {volume}
  {51}},\ \bibinfo {pages} {5600} (\bibinfo {year} {1995})},\ \Eprint
  {http://arxiv.org/abs/gr-qc/9409001} {arXiv:gr-qc/9409001} \BibitemShut
  {NoStop}%
\bibitem [{\citenamefont {Saltas}\ \emph {et~al.}(2014)\citenamefont {Saltas},
  \citenamefont {Sawicki}, \citenamefont {Amendola},\ and\ \citenamefont
  {Kunz}}]{Saltas:2014dha}%
  \BibitemOpen
  \bibfield  {author} {\bibinfo {author} {\bibfnamefont {I.~D.}\ \bibnamefont
  {Saltas}}, \bibinfo {author} {\bibfnamefont {I.}~\bibnamefont {Sawicki}},
  \bibinfo {author} {\bibfnamefont {L.}~\bibnamefont {Amendola}},  and \bibinfo
  {author} {\bibfnamefont {M.}~\bibnamefont {Kunz}},\ }\href {\doibase
  10.1103/PhysRevLett.113.191101} {\bibfield  {journal} {\bibinfo  {journal}
  {\emph {Phys. Rev. Lett.}}\ }\textbf {\bibinfo {volume} {113}},\ \bibinfo
  {pages} {191101} (\bibinfo {year} {2014})},\ \Eprint
  {http://arxiv.org/abs/1406.7139} {arXiv:1406.7139 [astro-ph.CO]} \BibitemShut
  {NoStop}%
\bibitem [{\citenamefont {Kimura}\ \emph {et~al.}(2012)\citenamefont {Kimura},
  \citenamefont {Kobayashi},\ and\ \citenamefont {Yamamoto}}]{Kimura:2011dc}%
  \BibitemOpen
  \bibfield  {author} {\bibinfo {author} {\bibfnamefont {R.}~\bibnamefont
  {Kimura}}, \bibinfo {author} {\bibfnamefont {T.}~\bibnamefont {Kobayashi}},
  and \bibinfo {author} {\bibfnamefont {K.}~\bibnamefont {Yamamoto}},\ }\href
  {\doibase 10.1103/PhysRevD.85.024023} {\bibfield  {journal} {\bibinfo
  {journal} {\emph {Phys. Rev. D}}\ }\textbf {\bibinfo {volume} {85}},\
  \bibinfo {pages} {024023} (\bibinfo {year} {2012})},\ \Eprint
  {http://arxiv.org/abs/1111.6749} {arXiv:1111.6749 [astro-ph.CO]} \BibitemShut
  {NoStop}%
\bibitem [{\citenamefont {Williams}\ \emph {et~al.}(2004)\citenamefont
  {Williams}, \citenamefont {Turyshev},\ and\ \citenamefont
  {Boggs}}]{Williams:2004qba}%
  \BibitemOpen
  \bibfield  {author} {\bibinfo {author} {\bibfnamefont {J.~G.}\ \bibnamefont
  {Williams}}, \bibinfo {author} {\bibfnamefont {S.~G.}\ \bibnamefont
  {Turyshev}},  and \bibinfo {author} {\bibfnamefont {D.~H.}\ \bibnamefont
  {Boggs}},\ }\href {\doibase 10.1103/PhysRevLett.93.261101} {\bibfield
  {journal} {\bibinfo  {journal} {\emph {Phys. Rev. Lett.}}\ }\textbf {\bibinfo
  {volume} {93}},\ \bibinfo {pages} {261101} (\bibinfo {year} {2004})},\
  \Eprint {http://arxiv.org/abs/gr-qc/0411113} {arXiv:gr-qc/0411113 [gr-qc]}
  \BibitemShut {NoStop}%
\end{thebibliography}%

\end{document}